\documentclass{aa}  

\usepackage{graphicx}
\usepackage{txfonts}
\usepackage[usenames,dvipsnames]{xcolor}
\usepackage{cancel}
\usepackage{ulem}
\usepackage{soul}
\usepackage{multirow}
\usepackage{comment}
\usepackage{booktabs}
\usepackage{amsmath}
\usepackage{stfloats}
\usepackage{siunitx}

\usepackage{natbib,twoopt}
\usepackage[breaklinks=true,colorlinks=true,linkcolor=NavyBlue!50!Emerald,citecolor=NavyBlue!50!Emerald,filecolor=blue,urlcolor=NavyBlue!50!Emerald]{hyperref} 
\bibpunct{(}{)}{;}{a}{}{,}             
\makeatletter
  \newcommandtwoopt{\citetads}[3][][]{\href{http://adsabs.harvard.edu/abs/#3}%
    {\def\hyper@linkstart##1##2{}%
     \let\hyper@linkend\@empty\citetalp[#1][#2]{#3}}}
  \newcommandtwoopt{\citepads}[3][][]{\href{http://adsabs.harvard.edu/abs/#3}%
    {\def\hyper@linkstart##1##2{}%
     \let\hyper@linkend\@empty\citep[#1][#2]{#3}}}
  \newcommandtwoopt{\citettads}[3][][]{\href{http://adsabs.harvard.edu/abs/#3}%
    {\def\hyper@linkstart##1##2{}%
     \let\hyper@linkend\@empty\citett[#1][#2]{#3}}}
  \newcommandtwoopt{\citetyearads}[3][][]%
    {\href{http://adsabs.harvard.edu/abs/#3}
    {\def\hyper@linkstart##1##2{}%
     \let\hyper@linkend\@empty\citetyear[#1][#2]{#3}}}
\makeatother

\begin{document} 

   \title{MAMBO - An empirical galaxy and AGN mock catalogue for the exploitation of future surveys}

   \author{X. López-López
          \inst{1,2}
          \and
          M. Bolzonella\inst{1}
          \and
          L. Pozzetti\inst{1}
          \and
          M. Salvato\inst{3,4}
          \and
          L. Bisigello\inst{5,6}
          \and
          A. Feltre\inst{7}
          \and
          I.E. López\inst{1}
          \and
          A. Viitanen\inst{8,9}
          \and
          V. Allevato\inst{10}
          \and
          A. Bongiorno\inst{8}
          \and
          G. Girelli\inst{11}
          \and
          J. Buchner\inst{3}
          \and
          S. Charlot\inst{12}
          \and
          F. Ricci\inst{13,8}
          \and
          C. Schreiber\inst{14}
          \and
          G. Zamorani\inst{1}
          }

   \institute{
    INAF-Osservatorio di Astrofisica e Scienza dello Spazio di Bologna, Via Piero Gobetti 93/3, 40129 Bologna, Italy 
   \and
    Dipartimento di Fisica e Astronomia, Universit\`a di Bologna, Via Gobetti 93/2, 40129 Bologna, Italy
    \and
    Max-Planck-Institut für extraterrestrische Physik, Giessenbachstr. 1, 85748 Garching, Germany
    \and
    Exzellenzcluster ORIGINS, Boltzmannstr. 2, 85748 Garching, Germany
    \and
    INAF, Istituto di Radioastronomia, Via Piero Gobetti 101, 40129 Bologna, Italy
    \and
     Dipartimento di Fisica e Astronomia "G. Galilei", Universit\`a di Padova, Via Marzolo 8, 35131 Padova, Italy
    \and
    INAF-Osservatorio Astrofisico di Arcetri, Largo E. Fermi 5, 50125, Firenze, Italy
    \and
    INAF-Osservatorio Astronomico di Roma, Via Frascati 33, 00078 Monteporzio Catone, Italy
    \and
    Department of Physics and Helsinki Institute of Physics, Gustaf Hällströmin katu 2, 00014 University of Helsinki, Finland
    \and
    INAF-Osservartorio Astronomico di Capodimonte, Via Moiariello 16, 80131 Napoli
    \and
    Independent researcher, Bologna, Italy
    \and
    Sorbonne Universit\'e, CNRS, UMR 7095, Institut d'Astrophysique de Paris, 98 bis bd Arago, 75014 Paris, France
    \and
    Dipartimento di Matematica e Fisica, Università degli Studi Roma Tre, via della Vasca Navale 84, I-00146 Roma, Italy
    \and
    IBEX Innovations, Sedgefield, Stockton-on-Tees, TS21 3FF, United Kingdom
    }
             
   \date{}

 
  \abstract
   {Current and future large surveys will produce unprecedented amounts of data. Realistic simulations have become essential for the design and development of these surveys, as well as for the interpretation of the results.}
   {We present MAMBO, a flexible and efficient workflow to build empirical galaxy and Active Galactic Nuclei (AGN) mock catalogues that reproduce the physical and observational properties of these sources.}
   {We start from simulated dark matter (DM) haloes, to preserve the link with the cosmic web, and we populate them with galaxies and AGN using abundance matching techniques. We follow an empirical methodology, using stellar mass functions (SMF), host galaxy AGN mass functions and AGN accretion rate distribution functions studied at different redshifts to assign, among other properties, stellar masses, the fraction of quenched galaxies, or the AGN activity (demography, obscuration, multiwavelength emission, etc.).}
   {As a proof test, we apply the method to a Millennium DM lightcone of 3.14 $\rm deg^2$ up to redshift $z=10$ and down to stellar masses $\mathcal{M} \gtrsim 10^{7.5} \, M_\odot$.  We show that the AGN population from the mock lightcone here presented reproduces with good accuracy various observables, such as state-of-the-art luminosity functions in the X-ray up to $z \sim 7$ and in the ultraviolet up to $z \sim 5$, optical/NIR colour-colour diagrams, and narrow emission line diagnostic diagrams. Finally, we demonstrate how this catalogue can be used to make useful predictions for large surveys. Using \textit{Euclid} as a case example, we compute, among other forecasts, the expected surface densities of galaxies and AGN detectable in the \textit{Euclid} $H_{\rm E}$ band. We find that \textit{Euclid} might observe (on $H_{\rm E}$ only) about $10^{7}$ and $8 \times 10^{7}$ Type 1 and 2 AGN respectively, and $2 \times 10^{9}$ galaxies at the end of its $\num{14679} \, \rm deg^2$ Wide survey, in good agreement with other published forecasts.
}
   {}

   \maketitle
%

\section{Introduction}
In the upcoming years, `full-sky' space and ground-based surveys, such as \textit{Euclid} \citep{EuclidSkyOverview}, MOONS \citep{MOONS.2020SPIE11447E..8DC}, Rubin/LSST \citep{Rubin.2019ApJ...873..111I} or eROSITA \citep{Merloni.2024A&A...682A..34M}, among others, will survey unprecedentedly large areas of the sky, gathering photometric and spectroscopic data for billions of galaxies and (at least) millions of Active Galactic Nuclei (AGN). Synthetic data reproducing observed properties of astrophysical sources have become essential to enhance the scientific return from this new data \citep[see e.g.][]{Georgakakis.2019MNRAS.487..275G, Comparat.2020OJAp....3E..13C, Allevato.2021ApJ...916...34A, Bisigello.2021A&A...651A..52B, Selwood.2024arXiv240518126E}. Before the start of observations, they are needed, e.g., to define the selection bias, test data analysis pipelines, and develop and calibrate models and parameters for future analyses. Synthetic data can also help in understanding the observations; in particular, it can be used to estimate incompleteness and biases and to refine and validate hypotheses on the basis of the simulations.

Different methods can be used to generate mock catalogues \citep[see][for a review]{Wechsler.2018ARA&A..56..435W}, ranging from physics-oriented models (hydrodynamic simulations or semi-analytic models) to data-oriented simulations (empirical models adopting observed scaling relations).
Many of these simulations are primarily focused on cosmological purposes, such as performing idealised tests on key cosmological probes like clustering and weak lensing \citep[e.g.][]{Korytov.2019ApJS..245...26K, Ishiyama.2021MNRAS.506.4210I, Hadzhiyska.2023MNRAS.525.4367H}. However, to generate reliable forecasts from these measurements, it is crucial to incorporate realistic galaxy and AGN properties to accurately account for observational selection and systematic effects.

Recently, large-scale simulations that accurately reproduce galaxy properties have been developed \citep[e.g.][]{Kovacs.2022OJAp....5E...1K, Wechsler.2022ApJ...931..145W,Gu.2024MNRAS.529.4015G,Castander.2024arXiv240513495E}, often adopting a semi-empirical or hybrid methodologies. These methods not only save computation time but also ensure that results align well with most observed scaling relations. In addition, thanks to their optimised computational efficiency, they can also be quickly adapted to implement new empirical relations following new discoveries. Our work is conducted within this framework: we have adopted a semi-empirical method to produce realistic samples of synthetic galaxies and AGN, a component often neglected in previous similar works, necessary to enhance the realism of our simulation.  At the same time, we preserve the link with the cosmic web traced by dark matter (DM) haloes: this connection is fundamental for deriving cosmological forecasts and linking the visible properties of galaxies and AGNs to the DM distribution.

The inclusion of AGN in these catalogues is of utmost importance; supermassive black holes (SMBHs) are believed to exist at the cores of most galaxies, and their masses are known to correlate closely with that of the host galaxy bulge, but also with their velocity dispersion and luminosity \citep{Richstone.1998Natur.395A..14R, Gebhardt.2000ApJ...539L..13G, Merritt2001ApJ...547..140M, Kormendy.2013ARA&A..51..511K}. Furthermore, galaxies hosting actively accreting SMBHs, i.e., AGN, frequently present characteristic observational features (both in photometric and spectroscopic observations) along the full electromagnetic spectrum. Reproducing these features in realistic empirical catalogues is essential.

Although the aim of this work is to produce mock NIR/optical catalogues, we decided to implement AGN starting from X-ray-selected samples due to the efficiency of this selection in identifying AGN across a wide range of redshifts and luminosities, including those with high levels of nuclear obscuration ($22 < \log N_{\rm H}/\rm cm^{-2} < 24$) that may be overlooked in other bands \citep[e.g.][]{Padovani.2017A&ARv..25....2P}. Since X-ray AGN photons are generated through non-thermal processes near the central black hole, they serve as an ideal tracer of AGN activity \citep{Mushotzky.1993ARA&A..31..717M, Brandt.2005ARA&A..43..827B}. Additionally, the X-ray band, especially for the brightest sources, experiences minimal contamination from the host galaxy, which predominantly emits at different wavelengths.

The structure of this paper is the following: in Sect.~\ref{sec:MAMBO}, we describe the methodology followed to create the galaxy mock catalogue which is used as a base for the consequent AGN mock. In Sect.~\ref{sec:AGN_on_MAMBO} we describe in detail the methodology followed to populate the galaxy mock with AGN. Then, in Sect.~\ref{sec:validation} we validate the catalogue by comparing its outputs with observations that were not used for its calibration, and finally in Sect.~\ref{sec:uses_future}, we show one example of how this catalogue can be used to make predictions for future surveys, focusing on \textit{Euclid} as a representative case.

Throughout this paper, a standard cosmology ($\Omega_{\rm m} = 0.3$, $\Omega_\Lambda = 0.7$ and $H_0 = 70\,{\rm km}\, {\rm s}^{-1}\, {\rm Mpc}^{-1}$) has been assumed. The stellar masses are given in units of solar masses for a Chabrier initial mass function \citep{Chabrier.2003PASP..115..763C}.

\section{Mock Galaxy catalogue: MAMBO} \label{sec:MAMBO}
\begin{figure}[h]
\centering
\includegraphics[width = 0.9\columnwidth]{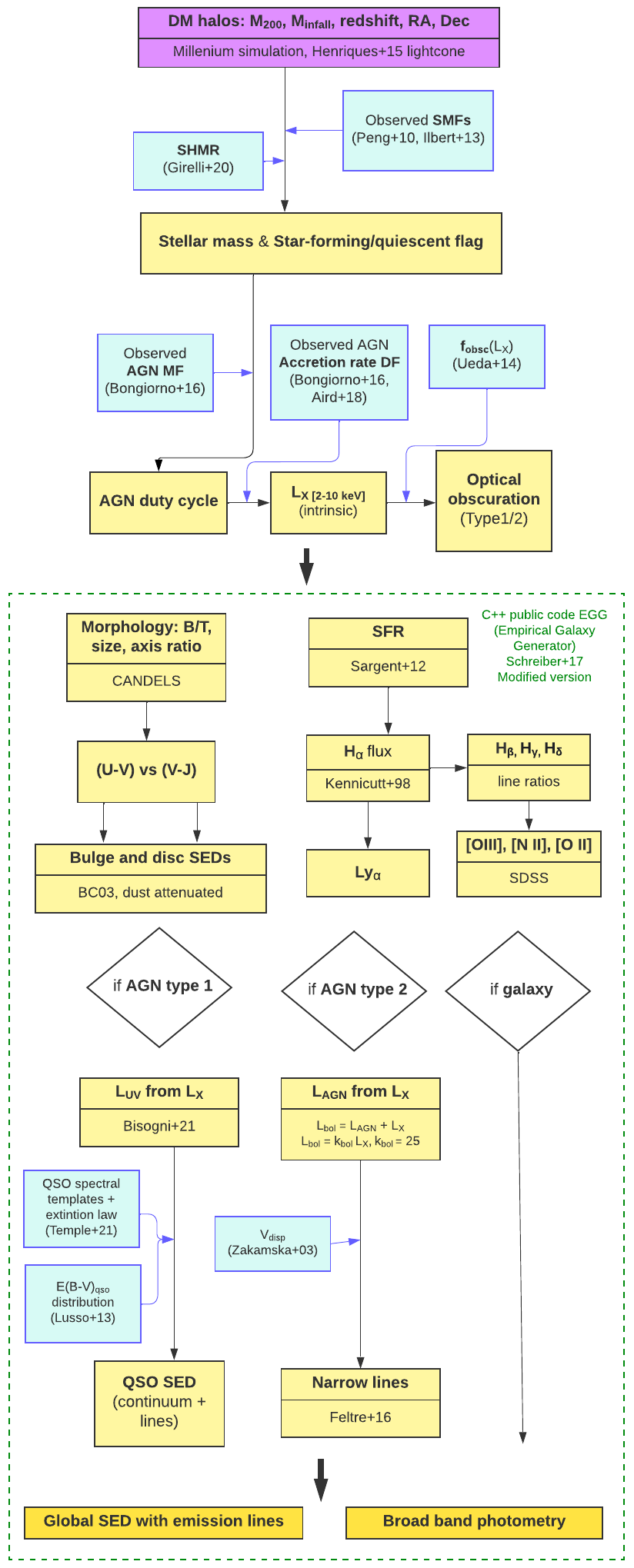}
\caption{Schematic view of the full workflow presented in this paper to create the galaxy and AGN mock catalogue. Yellow boxes represent the result of each step, while blue boxes show the necessary inputs. Steps within the green dotted box take place within the modified version of the public code EGG. A detailed description is given in Sects.~\ref{sec:MAMBO} and~\ref{sec:AGN_on_MAMBO}.}
\label{fig:workflow}
\end{figure}

MAMBO (Mocks with Abundance Matching in BOlogna) is a workflow designed to construct an empirical mock catalogue of galaxies which can reproduce with accuracy their physical properties and observables, such as rest-frame and observed magnitudes and spectral features. A detailed description of the methodology and a validation of the galaxy properties can be found in \citet{amsdottorato9820} and is briefly summarised in this section. In the rest of this paper, instead, we will focus our attention on the inclusion of AGN into this workflow. A schematic view of the steps explained in this section, as well as in Sect.~\ref{sec:AGN_on_MAMBO}, is presented in the flowchart of Fig.~\ref{fig:workflow}.

MAMBO takes in input few quantities from cosmological N-body DM simulations, i.e. the DM halo mass, expressed as $M_{200}$\footnote{Mass within the radius where the halo has an overdensity 200 times the critical density of the simulation.} for main haloes and $M_{\rm infall}$\footnote{Subhalo mass at the time it was accreted to the host halo; the infall mass is considered a better tracer of the potential well and then it correlates with galaxy properties such as the stellar mass \citep[e.g.][]{2006ApJ...647..201C,2010ApJ...710..903M}.} for subhaloes and orphans\footnote{Subhaloes that have lost all or part of their mass.}; the redshift $z$ of each halo and subhalo. The sky coordinates RA and Dec are not used as inputs, but are needed to reconstruct the cosmic web and derive clustering and environmental properties of different classes of objects. 

The results presented in the following are based on a lightcone built by \citet{2015MNRAS.451.2663H} using the Mock Map Facility (MoMaF, \citealt{momaf.10.1111/j.1365-2966.2005.09019.x}) on the Millennium Simulation \citep{Springel.2005Natur.435..629S}, namely lightcone number 23, which was chosen because it presents a mass function which is the closest to the mean of all the 25 available lightcones. The lightcone spans from $z = 0$ to $z = 10$ and contains DM haloes with a minimum mass $M_{\rm{halo}} \gtrsim 10^{10.24} \, M_\odot \, h^{-1}$, which corresponds to 20 DM particles, and it covers an area of $3.14\,{\rm deg}^2$. However, the method can be applied to any simulated catalogue of DM haloes and subhaloes.


In the first step, a galaxy with stellar mass $\mathcal{M}$ is assigned to each DM halo by means of a stellar-to-halo mass relation (SHMR). The SHMR was derived using a subhalo abundance matching technique, and calibrated on the Millennium lightcones by means of observed stellar mass functions (SMFs) on the SDSS \citep{SDSS.2000AJ....120.1579Y}, COSMOS \citep{UltraVISTA.2012A&A...544A.156M} and CANDELS \citep{Candels.2011ApJS..197...35G} fields. A detailed description of the SHMR can be found in \citet{Girelli_2020A&A...634A.135G} and \citet{amsdottorato9820}.

Additionally, every galaxy in the mock is classified as passive/quiescent (Q) or star-forming (SF) in a probabilistic way, following the relative ratio of the blue and red populations in observed SMFs. For this, we used the following SMFs: at $z \sim 0$, the SMF evaluated by \citet{Peng_2010ApJ...721..193P} on the SDSS survey and divided into passive and star-forming using the rest-frame ($U - B$) colour; at $0.2 < z < 4$, the SMF by \citet{Ilbert_refId0}, derived on the COSMOS field and classified into red/blue using the rest frame colour selection $(NUV - r)$ vs $(r - J)$ \citep{Ilbert.2010ApJ...709..644I}.

At $z \geq 4$, SMFs divided by SF/Q type are not available. Recent studies have tried to put constraints on this fraction \citep{Merlin17.1093/mnras/stx2385, Girelli19.refId0, Xie.2024ApJ...966L...2X}, but still with large uncertainties. For this reason, the star-forming fraction was extrapolated from the results at lower redshifts, arriving at a maximum of $f_{{\rm SF}} = 99\%$ at $z=6$, which is kept constant at higher redshifts. 
This choice is motivated by the fact that at that redshift, the Universe is supposed to be too young ($0.6\,{\rm Gyr}$ at $z=6$) to contain any considerable fraction of quiescent galaxies. Recently, new results from JWST data are starting to find small samples of quiescent galaxies at $3 < z < 6$ \citep{Carnall.2023MNRAS.520.3974C, Carnall.2024arXiv240502242C,Alberts.2023arXiv231212207A}, while \citet{2023arXiv230214155L} reported the finding of a quiescent galaxy at $z = 7.3$. We plan to revise these assumptions once sufficient data becomes available to accurately constrain the statistics of this population.

\begin{figure}[h]
\centering
\includegraphics[width = \columnwidth]{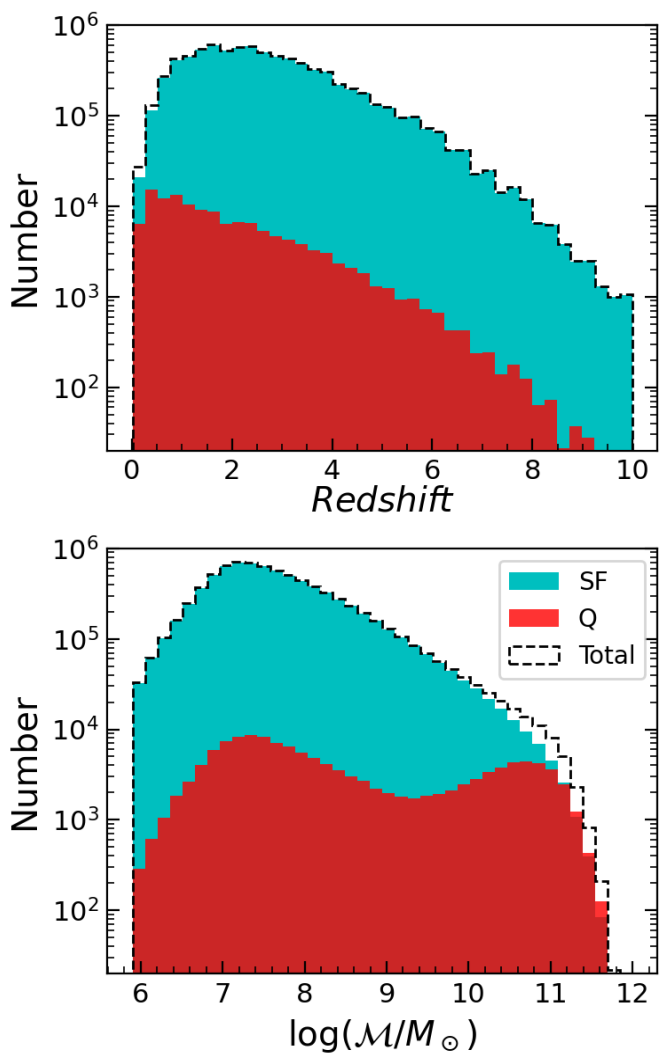}
\caption{Distribution of stellar mass and redshift of all the galaxies in the MAMBO lightcone used in this work, separated into the star forming (blue) and quiescent (red) populations. The dashed black line shows the total (SF+Q) galaxy population.}
\label{fig:SHMR_hists}
\end{figure}

We show in Fig.~\ref{fig:SHMR_hists} the stellar mass and redshift distributions resulting from applying the SHMR and the SF/Q classification process described above. As it is visible from this figure, the galaxy catalogue is complete at least down to  $\mathcal{M} \sim 10^{7.5} \, M_\odot$, which is a consequence of the mass completeness of the DM halo catalogue. Additionally, it can be seen from this figure that the stellar mass distribution of quiescent galaxies follows the shape of a double Schechter function, as is often the case at redshifts where it is possible to observe also low mass galaxies \citep{2009ApJ...707.1595D,Peng_2010ApJ...721..193P,Ilbert_refId0,2017A&A...605A..70D,2023A&A...677A.184W}.

As a final step, other physical properties, as well as the photometry and spectra of the galaxies in the catalogue are retrieved with a modified version of the public code EGG (Empirical Galaxy Generator, \citealp{EGG}). EGG is a C++ code designed to generate an empirical mock catalogue of galaxies with realistic physical properties (such as star formation rate, size, dust extinction, velocity dispersion and emission line luminosities), where every galaxy is treated as a two component system (bulge + disc). Additionally, the code produces the observed and rest-frame photometry in any desired band, and the redshifted (observer-frame) spectra from the UV to the submillimeter. The code has been calibrated purely by using empirical relations to produce realistic observable properties. In order to assign a SED (spectral energy distribution) to each galaxy, EGG selects an optical template from a prebuilt library of \citet{BC03.2003MNRAS.344.1000B} models, covering uniformly the observed part of the $UVJ$ plane \citep{Williams.2009ApJ...691.1879W}, which separates quiescent from star-forming galaxies.  Infrared SEDs are instead derived from a set of libraries \citep{Chary2001,Magdis2012,Schreiber2017} aimed at reproducing dust emission, characterised by the values of infrared luminosity of dust and PAHs (polycyclic aromatic hydrocarbons) at $\lambda = [8,1000]\,\mu$m, the dust temperature, and the ratio of IR to $8\,\mu$m luminosity \citep[IR8, ][]{Elbaz2011} for dust and PAH.

Afterwards, Gaussian emission lines with a given velocity dispersion\footnote{We modified the original recipe in EGG with a mass-dependent $\sigma_{\rm gas}$ from \citet{Bezanson2018ApJ...868L..36B}.} is assigned to the SEDs of the bulge and the disc components. EGG can take as an input the stellar mass $\mathcal{M}$, redshift $z$, and type (star-forming or quiescent) of each galaxy, or produce these quantities randomly extracting $z$ and $\mathcal{M}$ from the observed galaxy SMFs. We used EGG in the former configuration. More details about the way EGG works and about the modifications we did to the code are given in Sect.~\ref{sec:AGN_EGG}.

It is worth stressing that the full MAMBO workflow can be applied to any DM simulation, and, similarly, the method presented in Sect.~\ref{sec:AGN_on_MAMBO} to populate galaxies with AGN can be applied to any mock galaxy catalogue containing information about galaxy stellar mass, redshift and galaxy type.

\section{Adding AGN to the catalogue} \label{sec:AGN_on_MAMBO}
In this section, we present the methodology adopted to populate our galaxy catalogue with AGN, which uses a completely empirical methodology comprising the following steps: \textit{i)} we flag each galaxy from the lightcone as hosting an AGN or not, following a probabilistic method which depends on $\mathcal{M}$ and $z$ (Sect.~\ref{sec:AGN_frac}); \textit{ii)} every object flagged as AGN is assigned an intrinsic X-ray luminosity (Sect.~\ref{sec:X-ray}); \textit{iii)} we separate the AGN population into optically unobscured/obscured (Type 1/Type 2 respectively) by means of their intrinsic X-ray luminosity (Sect.~\ref{sec:obs}); \textit{iv)} we build the observed spectra and photometry of both Type 1 and Type 2 AGN with the help of photoionisation models of AGN narrow-line regions and parametric SED of typical QSO (that model both the continuum and emission lines; Sect.~\ref{sec:AGN_EGG}). A schematic view of this workflow is given in Fig.~\ref{fig:workflow}. 

\subsection{Specific accretion rate}
The specific accretion rate ($\lambda_{{\rm SAR}} \propto L_{\rm X}/\mathcal{M}$)\footnote{Throughout this paper we use $L_{\rm X}$ and $F_{\rm X}$ to refer, respectively, to the intrinsic luminosity and observed flux in the hard ([$2-10 \rm keV$]) X-ray band.} is a directly measurable quantity which has been extensively used in the literature \citep{Bongiorno2012MNRAS.427.3103B, Aird2012ApJ...746...90A, Georgakakis.2014MNRAS.440..339G} as a proxy for the rate of accretion onto the SMBH ($\dot{M}_{{\rm BH}}$) relative to the stellar mass of the host galaxy. Furthermore, if a bolometric correction and a $M_{{\rm BH}}/\mathcal{M}$ scaling relation are assumed, the specific accretion rate can also be regarded as a proxy for the Eddington ratio of the SMBH, $\lambda_{{\rm Edd}} = L_{{\rm bol}}/M_{{\rm BH}}$ \citep{Bongiorno16,Aird18}. 

In the following subsections, we use two specific accretion rate distribution functions (SARDFs) with different definitions. On one hand, \citet{Bongiorno16} defines the specific accretion rate as $\lambda_{{\rm SAR}} = L_{\rm X}/\mathcal{M}$. On the other hand, \citet{Aird18} defines the specific black hole accretion rate ($\lambda_{{\rm sBHAR}}$) as the dimensionless quantity:

\begin{equation}
    \lambda_{{\rm sBHAR}}=\frac{k_{{\rm bol}} L_{\rm X}}{1.3 \times 10^{38} {\rm erg\,s^{-1}} \times 0.002 \frac{\mathcal{M}}{M_{\odot}}},
\label{eq:sbhar}
\end{equation}

\noindent where $k_{{\rm bol}}$ is a bolometric correction factor ($L_{{\rm bol}} = k_{\rm bol}\,L_{\rm X}$), assumed to have a constant value of $k_{{\rm bol}} = 25$. This definition also assumes a constant scaling relation $M_{{\rm BH}} = 0.002\mathcal{M}$ (\citealt{Mbh.Mgal.2003ApJ...589L..21M}, assuming also $\mathcal{M} \approx M_{\rm bul}$). Under these assumptions, $\lambda_{{\rm sBHAR}} = \lambda_{{\rm Edd}}$, and therefore an AGN accreting at $1\%$ of the Eddington limit would have $\lambda_{{\rm sBHAR}} \sim 10^{-2}$. It is possible to convert between the two definitions of the specific accretion rate as $\lambda_{{\rm SAR}} \approx 10^{34}\,\lambda_{{\rm sBHAR}}$. Therefore, with the same assumptions, the $1\%$ of the Eddington limit corresponds to $\lambda_{{\rm SAR}} = 10^{32}\, {\rm erg\,s^{-1}} M_\odot^{-1}$. We note that both in \citet{Bongiorno16} and in \citet{Aird18}, this was set as the lower limit to define a galaxy as hosting an AGN.

\subsection{Fraction of AGN}\label{sec:AGN_frac}

In order to derive the probability of a galaxy with a given stellar mass $\mathcal{M}$ and at a given redshift $z$ to be hosting an active nucleus, $p\left({\rm AGN} \mid \mathcal{M}, z\right)$, we make use of the AGN host galaxy mass function (HGMF) derived by \citet[][hereafter B16]{Bongiorno16}. In B16 the authors studied a sample of $877$ hard ($2-10\,{\rm keV}$) X-ray selected AGN from the XMM-COSMOS point-like source catalogue \citep{Hasinger_XMM_COSMOS,Cappelluti_XMM_COSMOS,Brusa.2010ApJ...716..348B,Bongiorno2012MNRAS.427.3103B} in the redshift range $0.3 < z < 2.5$ and with a limiting flux of $F_{\rm X}\sim 3 \times 10^{-15}\, \rm erg\,s^{-1}\,cm^{-2}$. This sample was also selected with a stellar mass limit of $\mathcal{M} > 10^{9.5} \, M_\odot$ and specific accretion rate $\lambda_{{\rm SAR}} > 10^{32}\, \rm erg\,s^{-1} M_\odot^{-1}$.

To derive the AGN HGMF and the SARDF, B16 corrected for the stellar mass incompleteness of the sample down to the above-mentioned limit in $\lambda_{\rm SAR}$. Additionally, B16 also accounted for the incompleteness due to the sources that were missed in the sample because of their high column density $N_{\rm H}$. This was done considering column density values in the range $20 < \log N_{\rm H}/\rm cm^{-2} < 24$, and therefore not including Compton-thick AGN, i.e., heavily obscured objects with $\log N_{\rm H}/\rm cm^{-2} > 24$. Although a significant fraction of AGN are expected to be CTK sources, their exact fraction is still a matter of debate (e.g. \citealp{Buchner2015ApJ...802...89B} found a constant fraction of $f_{\rm CTK} \sim 35\%$, independent of redshift and accretion luminosity, while \citealp{Pouliasis.2024A&A...685A..97P} found a much lower fraction of $f_{\rm CTK} \sim 17\%$ for $3 \lesssim z \lesssim 6$ AGN). Furthermore, the redshift and luminosity-dependence of this fraction is yet not fully understood \citep[e.g.][]{C.Ricci.2017Natur.549..488R}. Therefore, the inclusion of these sources is left to future work.

It is worth noting that the choice of a minimum value of $\lambda_{{\rm SAR}}$ sets the definition of AGN used in this work. In general, different criteria can be used to select AGN based on their X-ray emission. A common method is to select sources emitting above a threshold intrinsic X-ray luminosity (generally $L_{\rm X} > 10^{42} \,{\rm erg}\,{\rm s}^{-1}$, e.g. \citealp{Brandt_alexander_2015A&ARv..23....1B}). AGN can also be selected according to the relative X-ray emission from non-nuclear mechanisms (such as X-ray binary stars and hot gas emission) with respect to the total $L_{\rm X}$ of the galaxy \citep[e.g.][]{Birchall_20_10.1093/mnras/staa040}. It is well known that selecting AGN with these different criteria can lead to significant differences in their observed fraction \citep[see e.g.][]{Birchall_2022MNRAS.510.4556B}, which should be taken into consideration when using this catalogue.

In B16 the authors derived the HGMF by jointly fitting the SMF and the SARDF (see Sect.~\ref{sec:X-ray}), with the X-ray luminosity function (XLF) as an additional constraint. For this purpose, they used a maximum likelihood method to determine the HGMF and the SARDF as a bivariate distribution function of stellar mass and specific accretion rate, $\Psi(\mathcal{M},\lambda_{{\rm SAR}},z)$. As a result of this approach, the HGMF cannot be expressed as a simple analytic function, but instead, the authors provide an analytic approximation by performing a least-squares fit to the HGMF with a Schechter \citep{Schechter1976ApJ...203..297S} function:

\begin{equation}
    \Phi(\mathcal{M})\: {\rm d}\mathcal{M} = \Phi^{\star} \left(\frac{\mathcal{M}}{\mathcal{M}^{\star}}\right)^\alpha {\rm exp}{\left(-\frac{\mathcal{M}}{\mathcal{M}^{\star}}\right)}\: {\rm d}\left(\frac{\mathcal{M}}{\mathcal{M}^{\star}}\right),
\label{eq:schechter}
\end{equation}

\noindent evaluated at the centre of 3 redshift bins ($0.3 < z < 0.8$, $0.8 < z < 1.5$ and $1.5 < z < 2.5$). The best-fit parameters of this fit are given in Table~2 of B16.\footnote{B16 used a different definition for the Schechter function, therefore to use the values given in their Table~2 one must remove the term ${\rm d}\left(\frac{\mathcal{M}}{\mathcal{M}^{\star}}\right)$ from Eq.~\ref{eq:schechter}.} 

The original slope ($\alpha$) from the Schechter function in the first two redshift bins would produce an unrealistic overestimate of low mass AGN when extrapolating this model to $\mathcal{M} < 10^{9.5} M_\odot$, which is the stellar mass limit in the sample of B16, while our catalogue spans to lower stellar masses as visible in Fig.~\ref{fig:SHMR_hists} (see Appendix~\ref{sec:appendix1} for an example using the original slopes). Therefore, we re-derived these quantities by fitting the $1/V_{{\rm max}}$ points shown in B16 with a Schechter function using Eq.~\ref{eq:schechter}, obtaining $\alpha = -0.25 \pm 0.09$ for $0.3 < z < 0.8$, and $\alpha = -0.19 \pm 0.11$ for $0.8 < z < 1.5$. For the highest redshift bin, our fit was compatible with the value shown in Table~2 of B16, therefore we didn't modify it.

We then calculate $p\left({\rm AGN} \mid \mathcal{M}, z\right)$ as the ratio between the SMF of the galaxies of the MAMBO lightcone and the HGMF from B16. Because the MAMBO lightcone covers a wider range of redshifts ($0 < z < 10$), we interpolate and extrapolate $p\left({\rm AGN} \mid \mathcal{M}, z\right)$ from the centre of each redshift bin. For this, we assume a minimum fraction $p\left({\rm AGN} \mid  z = 0\right) = 0.01$ at all $\mathcal{M}$ when extrapolating at $z < 0.55$, while we maintain a constant fraction when extrapolating at $z>2$. Afterwards, every galaxy is statistically assigned as hosting an AGN or not with a Bernoulli trial proportional to $p$(AGN) (i.e., by comparing $p$(AGN) to a random number extracted from a uniform probability distribution from 0 to 1).

Although the choice of a minimal fraction of $1\%$ at $z\sim 0$ for all masses is a rough approximation, recent studies of the local Universe motivate this assumption. For example, \citet{Birchall_2022MNRAS.510.4556B} studied 917 X-ray selected AGN (found as XMM counterparts of 25,949 SDSS galaxies) with $z\leq 0.33$, which corresponds to a global AGN fraction of $3.5\%$. Instead, when selecting AGN by means of their accretion rate they found a fraction of about $1\%$, constant over stellar masses of $8 < \log(\mathcal{M}/M_\odot) < 12$, and increasing from $1\%$ to about $10\%$ with redshift. In \citet{2022MNRAS.510.4909W} the authors studied 213 \textit{Chandra} X-ray counterparts of 280 nearby ($< 120 {\rm Mpc}$) galaxies from the Palomar sample and classified 14 ($6.6\%$) of them as Seyferts, while only 4 ($1.9\%$) of them have $L_{\textrm{X}} > 10^{42}$. Similarly, \citet{Osorio_2023MNRAS.522.5788O} studied 138 \textit{Chandra} X-ray counterparts of the CALIFA sample, with a wide range of stellar masses and $z<0.1$ and found an AGN fraction of $5\%$. 
At $z>2$ we chose to have a constant AGN fraction: despite being a rough approximation, with this choice we are able to reproduce the observed X-ray luminosity functions up to $z=7$, as shown in Fig.~\ref{fig:xlf}.

\begin{figure}[h]
\centering
\includegraphics[width = \columnwidth]{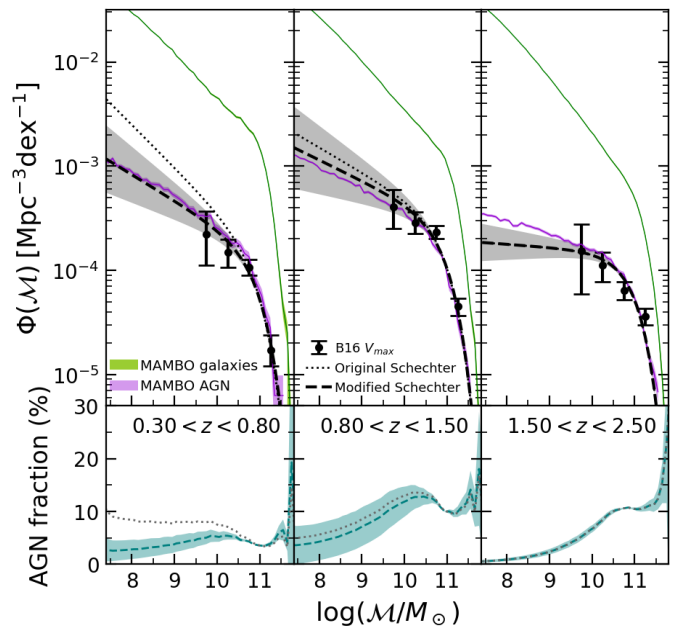}
\caption{ \textit{Upper panels:} Stellar mass function of galaxies and AGN. The green and purple shaded areas correspond to the galaxies and AGN from the MAMBO lightcone respectively. We also show the best-fit Schechter function to the host galaxy mass function (HGMF) of AGN from \citet{Bongiorno16} with a black dotted line, and the modified Schechter fit used in this work with a black dashed line. The shaded grey area shows the uncertainty in $\alpha$. Black dots with errorbars show the AGN mass function computed using the $V_{max}$ method, derived in B16. The three panels correspond to the three redshift bins studied in B16. \textit{Lower panels:} AGN fraction derived as the ratio of the modified Schechter function and the galaxy mass function from MAMBO (blue dashed line), and the same but using the original Schechter function from B16 (black dotted line).}\label{fig:agn_frac}
\end{figure}

\begin{figure}[h]
\centering
\includegraphics[width = \columnwidth]{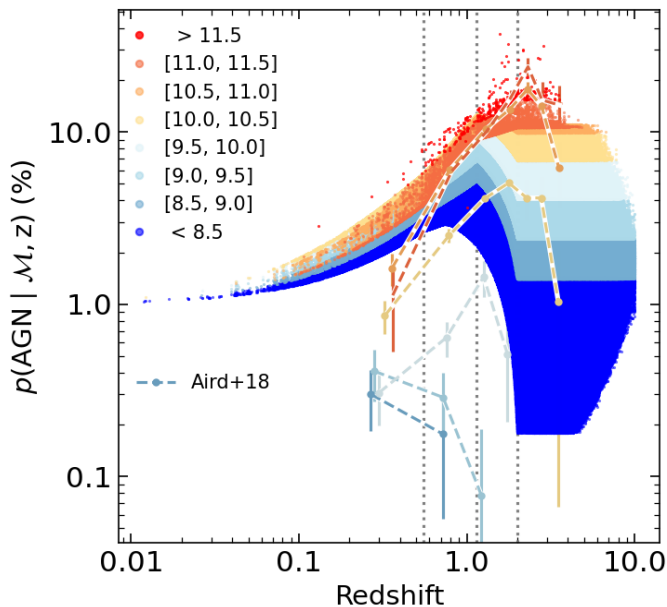}
\caption{Probability of a galaxy hosting an AGN as a function of redshift. Each point corresponds to a galaxy in the MAMBO lightcone, colour-coded in different $\mathcal{M}$ bins indicated in the legend. For comparison, the points with errorbars connected by dashed lines show the AGN fraction derived in A18, also in different mass bins. Vertical dotted lines mark the centre of the three redshift bins studied in B16, namely $z = 0.55, 1.15, 2.0$.}
\label{fig:agn_frac_mam}
\end{figure}

In the upper panels of Fig.~\ref{fig:agn_frac} we show the best-fit Schechter function of the AGN HGMF (both the original from B16 and our modified fit) together with the SMF of the galaxies of the MAMBO lightcone and the AGN mass function of our catalogue, derived with the methodology described above. We also show the AGN HGMF computed using the $V_{\rm max}$ method, derived in B16, as an additional consistency check to the Schechter model. As expected by construction, the AGN mass functions of MAMBO reproduce those of B16. An exception is the highest redshift bin, where the low-mass end of the HGMF of MAMBO is higher, but still compatible with the $V_{\rm max}$ estimate, due to the fact that we kept a constant $p\left({\rm AGN} \mid \mathcal{M}, z\right)$ when extrapolating at $z>2$ (see Fig.~\ref{fig:agn_frac_mam}). In the lower panels of Fig.~\ref{fig:agn_frac} we show the fraction of AGN over the galaxy population as a function of stellar mass and at a given redshift bin, $p\left(\rm AGN \mid \mathcal{M}, \langle z\rangle\right)$. As expected, this fraction increases with increasing stellar mass of the host galaxy. 

We show in Fig.~\ref{fig:agn_frac_mam} the probability of a given galaxy to be hosting an AGN as a function of redshift and in different mass bins for the full lightcone, after performing the interpolations and extrapolations described above. Each dot in this figure corresponds to a different galaxy, which is flagged as AGN (or not) in a statistical way depending on this probability. This probability, at all masses, increases with redshift reaching a maximum around $z\simeq 1$ and decreasing at higher redshifts. For comparison, Fig.~\ref{fig:agn_frac_mam} shows also the fraction of AGN as derived in \citet[hereafter A18]{Aird18}, where they derived the duty cycle of a sample of NIR selected galaxies matched with X-ray data. We note that the fractions differ significantly, being higher in this work, in particular at $\mathcal{M}<10^{10}\,M_\odot$. We show in Appendix~\ref{sec:appendix1} that using the AGN fractions from A18 as an input in our workflow, produces an X-ray luminosity function which tends to be underestimated when comparing it to the observed ones, especially for $z \lesssim 1$ and $z \gtrsim 3$.

\subsection{X-ray luminosity} \label{sec:X-ray}
In order to assign to each object flagged as AGN an intrinsic luminosity in the hard ([$2-10 \textrm{ keV}$]) X-ray band, we first assign a specific accretion rate. At $0 < z < 2$ we make use of the SARDF derived in B16. As explained in Sect.~\ref{sec:AGN_frac}, B16 derived the SARDF and the HGMF simultaneously as a bivariate distribution function of $\mathcal{M}$ and $\lambda_{{\rm SAR}}$, that is, $\Psi(\mathcal{M},\lambda_{{\rm SAR}},z)$. As was the case with the HGMF, this implies that the SARDF cannot be expressed as a simple analytic function of $\mathcal{M}$ and $z$. Instead, we used an analytic approximation of the SARDF (evaluated at the centre of three redshift bins) described as a double power-law (DPL) of the form:

\begin{equation}
    \Phi(\lambda_{{\rm SAR}}, \mathcal{M}) = \frac{\Phi^\star_\lambda}{\left(\frac{\lambda_{{\rm SAR}}}{\lambda^\star_{{\rm SAR}}\left(\mathcal{M}\right)}\right)^{-\gamma_1} + \left(\frac{\lambda_{{\rm SAR}}}{\lambda^\star_{{\rm SAR}}\left(\mathcal{M}\right)}\right)^{-\gamma_2}}\,,
    \label{eq:powerlaw}
\end{equation}

\noindent where the mass dependence is given by $\log \lambda^\star_{{\rm SAR}}\left(\mathcal{M}\right) = \log \lambda^\star_{{\rm SAR},0} + k_\lambda(\log \mathcal{M} - \log \mathcal{M}_{0})$, where $\log \lambda^\star_{{\rm SAR},0} = 33.8$, $\log \mathcal{M}_{0} = 11$ and $k_\lambda = 0.58$. The best-fit values of the normalisation $\Phi^\star_\lambda$ and slopes $\gamma_1$, $\gamma_2$ of the DPL evaluated at the centre of the three redshift bins are given in Table 3 of B16.

\begin{figure}[h]
\centering
\includegraphics[width = \columnwidth]{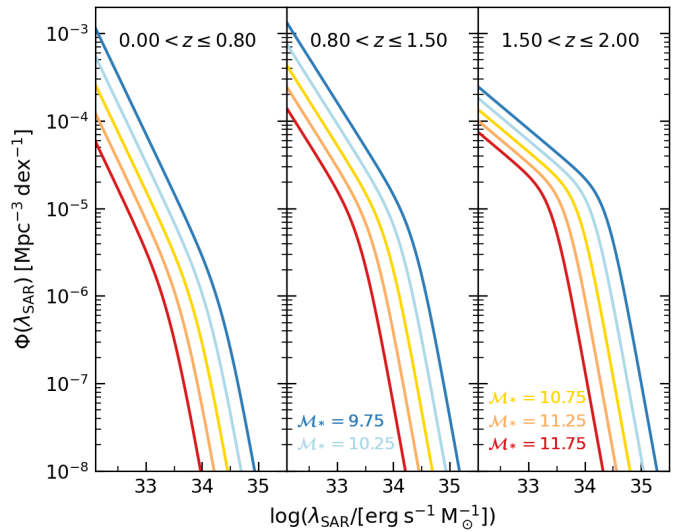}
\caption{Specific accretion rate distribution function (SARDF) computed at different values of $\mathcal{M}$ using Eq.~\ref{eq:powerlaw}, in the three redshift bins from B16.}
\label{fig:SARDF}
\end{figure}

Using Eq.~\ref{eq:powerlaw} we compute the SARDF at 5 values of $\mathcal{M}$, from $\log(\mathcal{M}/M_\odot) = 9.75$ to  $\log(\mathcal{M}/M_\odot) = 11.75$ in steps of 0.5\,dex in $\mathcal{M}$ (Fig.~\ref{fig:SARDF}). We then divide the AGN in our lightcone into 5 mass bins, each of them centred at one of the above-mentioned $\mathcal{M}$ and all of them of width 0.5\,dex in $\mathcal{M}$, except the lowest mass bin, extending up to the lowest $\mathcal{M}$ in our lightcone. Then, we normalise the $\lambda_{{\rm SAR}}$ distribution functions by dividing them by the integrated density of AGN (${\rm Mpc}^{-3}$) in each bin of $\mathcal{M}$ and $z$, therefore converting these distributions into probability distributions (PDF). We assign to each AGN a value of $\lambda_{{\rm SAR}}$ by randomly extracting values from the corresponding PDF at each $\mathcal{M}$ and $z$ bin.

At higher redshifts ($z > 2$), instead,  we make use of the accretion rate distributions from A18, who studied a sample of 1,797 X-ray \textit{Chandra} counterparts of 126,971 NIR-selected galaxies up to $z\sim 4$ and with stellar and $8.5 < \log(\mathcal{M}/M_\odot) < 11.5$.
 
Using a methodology similar to that used for the B16 accretion rate distributions previously described, we constructed the cumulative distribution function relative to each $p\left(\log\lambda_{{\rm sBHAR}} \mid \mathcal{M}, z\right)$ distributions in the range $-2 < \log\lambda_{{\rm sBHAR}} < 1$ and $2 < z < 4$, and use it to assign a value of $\lambda_{{\rm sBHAR}}$ to each object previously labeled as AGN at $z > 2$. Since A18 calculated the $\lambda_{{\rm sBHAR}}$ distributions separately for star-forming and quiescent galaxies, we use both sets of distributions for the galaxies in our catalogue that are split into these two classes. 

We also explored the possibility of using the accretion rate distributions from A18 to assign $L_{\rm X}$ at all redshifts, but we decided to use the methodology presented in this section since it produces an XLF which is in better agreement with the observed ones (see Appendix~\ref{sec:appendix1}).

\begin{figure}[h]
\centering
\includegraphics[width = \columnwidth]{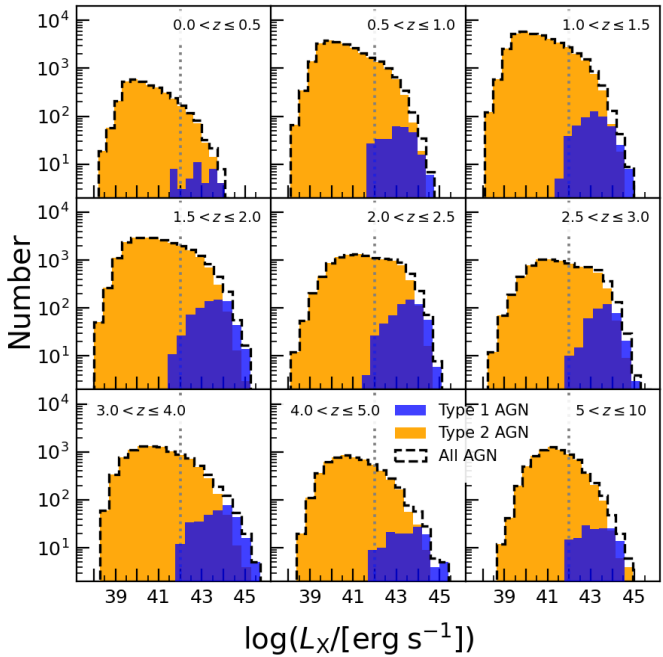}
\caption{Distribution of intrinsic hard-band X-ray luminosity ($L_{\rm X}$) of all the AGN in the lightcone, separated into Type 1 (blue) and Type 2 (orange) AGN, in different redshift bins. We also show with dashed and solid lines the subpopulations of Type 1 and 2 in star forming and quiescent galaxies. The vertical dotted line marks $L_{\rm X} = 10^{42} \,{\rm erg}\,{\rm s}^{-1}$, the threshold typically applied to separate X-ray emission from AGN or from other origins.}
\label{fig:Lx_hist}
\end{figure}

After assigning every AGN in the lightcone with a specific accretion rate, we convert this into intrinsic X-ray luminosity (in the $2-10\,{\rm keV}$ band). We show in Fig.~\ref{fig:Lx_hist} the histogram of the assigned $L_{\rm X}$ in different redshift bins for the entire population of AGN. We note that, as visible in Fig~\ref{fig:Lx_hist}, a large fraction of AGN have luminosities $L_{\rm X} < 10^{42} \,{\rm erg}\,{\rm s}^{-1}$, that is, below a classical threshold applied to define an X-ray emitter as an AGN. The reason for this is twofold: on one hand, as previously mentioned, we started from a sample of AGN selected above a given accretion rate limit and not above a $L_{\rm X}$ limit. On the other hand, the galaxy mock probes low stellar masses ($\mathcal{M} \sim 10^7 \, M_\odot$), which translates into low X-ray luminosities. 

\begin{figure}[h]
\centering
\includegraphics[width = \columnwidth]{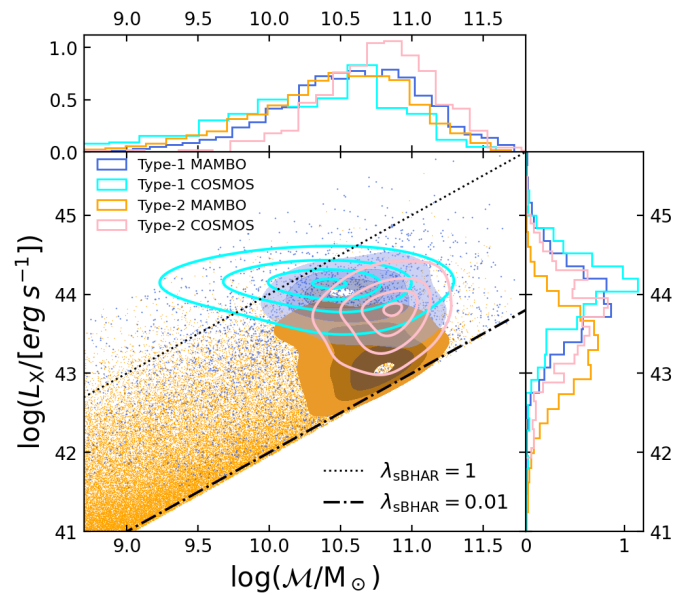}
\caption{AGN from our lightcone scattered in the $L_{\rm X} - \mathcal{M}$ plane, separated into Type 1 (blue) and Type 2 (orange) AGN. The points in the central panel show the distribution of all the sources, while the solid contours and the histograms from the upper and right panel correspond to sources selected above $F_{\rm X} = 1.9 \times 10^{-15}\, \rm erg\,s^{-1}\,cm^{-2}$. We show also the distribution of sources from the Chandra COSMOS Legacy Spectral Survey \citep{Marchesi.2016ApJ...830..100M}, with the same cut in $F_{\rm X}$, and separated into Type 1 (cyan) and Type 2 (pink) AGN. The contour levels represent iso-density lines, corresponding to the 50th, 75th, 90th, and 99th percentiles of the distribution. The dash-dotted and dotted black lines mark the locus where $\lambda_{{\rm sBHAR}} = 0.01$ and 1 respectively, which assuming a mean bolometric correction $k_{{\rm bol}} = 25$  and a constant mass ratio of black hole to host galaxy $M_{{\rm BH}} \approx 0.002\mathcal{M}$ correspond approximately to 1 and 100 percent of the Eddington limit respectively.} 
\label{fig:LxvsM}
\end{figure}

In Fig.~\ref{fig:LxvsM} we show the distribution of the AGN in our mock in the $L_{\rm X} - \mathcal{M}$ plane. As expected by construction, all the AGN have accretion rates above $\lambda_{{\rm sBHAR}} > -0.01$ ($\equiv \lambda_{{\rm SAR}} > 10^{32}\,\rm erg\,s^{-1} M_\odot^{-1}$), and only a small fraction of them are above the Eddington limit $\lambda_{{\rm sBHAR}} = 1$.


\subsection{Obscuration model} \label{sec:obs}

 The observed spectra of AGN in different bands can vary significantly depending on the level of obscuration of the nucleus by its surrounding material along the line of sight. In this context, AGN are generally separated into two main sub-classes, namely Type 1 and Type 2 AGN (unobscured and obscured respectively). In the UV/optical/NIR bands, Type 1 AGN are characterised by the presence of broad optical emission lines (full width at half-maximum $\gtrsim 1000\textrm{ km s}^{-1}$) and a SED with a bluer continuum, while Type 2 AGN typically don't present these characteristics. In reality, AGN display a much wider observational variety which allows us to classify them in many other sub-classes \citep[e.g.][]{Antonucci.1993ARA&A..31..473A, Urry.1995PASP..107..803U, Spinoglio.2021IAUS..356...29S}, but such classification is beyond the scope of this work.

 In order to separate the AGN of our catalogue into optically obscured and unobscured, we make use of the results presented in \citet{Merloni14.2014MNRAS.437.3550M}. In this study, the authors studied a sample of 1310 AGN selected from the XMM-COSMOS point-like source catalogue \citep{Hasinger_XMM_COSMOS,Cappelluti_XMM_COSMOS} with a limiting rest-frame X-ray flux of $F_{\rm X} = 2 \times 10^{-15} \textrm{ erg\,cm}^{-2}\,\textrm{s}^{-1}$ in the redshift range $0.3 < z < 3.5$. For this sample, the authors studied the luminosity dependence of the (optically) obscured fraction of AGN, and found a relation, almost redshift indipendent, of the form

 \begin{equation}
    F_{{\rm obs}}=A+\frac{1}{\pi} \operatorname{atan}\left(\frac{l_0-\log L_{\rm X}}{\sigma_x}\right),
    \label{eq:obs_frac}
\end{equation}

\noindent  where $F_{{\rm obs}}$ is the fraction of Type 2 AGN with respect to the total population, $L_{\rm X}$ is the intrinsic X-ray luminosity in the hard band, and the best-fitting parameters are $A=0.56, l_0=43.89 \text { and } \sigma_x=0.46$. For this work, we assumed this relation to be valid for the full redshift range $0 < z < 10$ without any redshift evolution. Using this relation, we assign to every AGN a probability of being optically obscured depending on their $L_{\rm X}$, and statistically separate every AGN in Type 1 and Type 2.

Additionally, the X-ray emission from AGN (especially in the soft band) can be heavily obscured by gas and dust surrounding the SMBH. To quantify this effect, we assign to each AGN a value of absorption column density ($N_{\rm H}$), following the absorption function presented in \citet{Ueda2014ApJ...786..104U}, within the range $20 < \log N_{\rm H}/\rm cm^{-2} < 24$. The choice of this range is consistent with the fact that, as stated in Sect.~\ref{sec:AGN_frac}, Compton-thick AGN ($\log N_{\rm H}/\rm cm^{-2} > 24$) are not included in our model. Following Section 3 of \citet{Ueda2014ApJ...786..104U}, we assign $N_{\rm H}$ in a probabilistic way as a function of $L_{\rm X}$ and redshift. This allows us to classify the AGN in our catalogue as X-ray Type 1 and Type 2, where Type 1 X-ray AGN are defined as an object with $\log N_{\rm H}/\rm cm^{-2} < 22$, and vice-versa.

It is worth noting that in nature these two classifications, although on average correlate with each other, do not match perfectly, i.e., some objects present optical obscuration but not X-ray obscuration and vice-versa \citep[e.g.][]{Marchesi.2016ApJ...830..100M, Merloni14.2014MNRAS.437.3550M}. Because the aim of this work is to produce a catalogue to be used in optical/NIR surveys (e.g. \textit{Euclid}, MOONS), in the following sections we will focus on Type 1 AGN defined as optically unobscured sources, regardless of their X-ray obscuration.

\begin{figure}[h]
\centering
\includegraphics[width = \columnwidth]{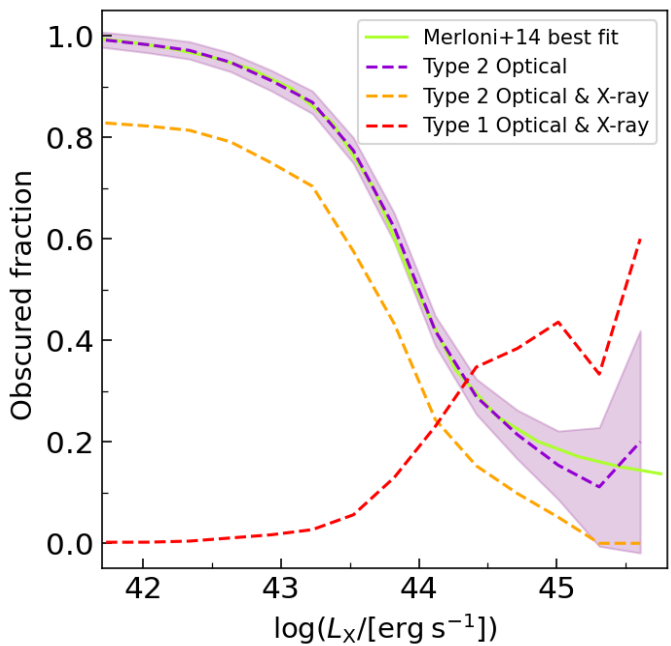}
\caption{Fraction of optically obscured (Type 2) AGN as a function of intrinsic X-ray luminosity. The green solid line shows the relation used to separate AGN (Eq.~\ref{eq:obs_frac}, originally from \citealp{Merloni14.2014MNRAS.437.3550M}) into optically obscured/unobscured, while the purple dashed line shows the actual fraction of Type 2 AGN in our catalogue (with the Poissonian uncertainty shown by the shaded area). The other two dashed lines give information about the X-ray obscuration (where Type 1 X-ray AGN are defined as objects with $\log N_{\rm H}/\rm cm^{-2} < 22$ and vice-versa): the orange dashed line shows the fraction of objects classified as Type 2 both in optical and the X-ray, while the red dashed line represents the fraction of all the AGN which are Type 1 in both bands at the same time.}
\label{fig:obsc_frac}
\end{figure}

 In Fig.~\ref{fig:Lx_hist} we show the $L_{\rm X}$ distribution separated into Type 1 and Type 2 AGN. It is visible from this figure that most of the AGN with $L_{\rm X} < 10^{42} \,{\rm erg}\,{\rm s}^{-1}$ are classified as Type 2 AGN in our catalogue. 
 
 In Fig.~\ref{fig:LxvsM} we show also the distribution of sources from the Chandra COSMOS Legacy Spectral Survey \citep{Marchesi.2016ApJ...830..100M} in the $L_{\rm X} - \mathcal{M}$ plane, selected above a minimum X-ray flux $F_{\rm X} = 1.9 \times 10^{-15}\, \rm erg\,s^{-1}\,cm^{-2}$, which is the flux limit reported in \citet{Marchesi.2016ApJ...830..100M}. We applied the same flux cut to the AGN from our catalogue (contours in Fig.~\ref{fig:LxvsM}) in order to compare with the COSMOS data. With this flux cut, our AGN and the COSMOS AGN are limited to $z \lesssim 3.5$. We observe a reasonable agreement between the MAMBO and the COSMOS distributions, although MAMBO Type 1 AGN seem to cover a narrower range in stellar mass than the COSMOS ones, and MAMBO Type 2 AGN have on average lower X-ray luminosities than their COSMOS counterparts.

 In Fig.~\ref{fig:obsc_frac} we show the fraction of optically obscured AGN as a function of $L_{\rm X}$ for the presented lightcone, and compare it with the calibration used to derive it, that is, Eq.~\ref{eq:obs_frac}. We also show the fraction of objects that are obscured both in optical and X-ray, and those that are unobscured in both bands.

\subsection{AGN emitted spectra}\label{sec:AGN_EGG}
After every galaxy in the catalogue has been classified as either hosting an AGN or not, and AGN have been characterised in terms of their X-ray luminosity and optical obscuration, we employ the publicly available code EGG (Empirical Galaxy Generator, \citealt{EGG}) to assign the rest of physical properties and observables (e.g. bulge/disc ratio, dust attenuation, etc.). Additionally, EGG allows us to generate the photometry of every object in any desired band, as well as the complete SED from the UV to the submillimeter.

In EGG, each galaxy is represented as a two-component system composed of a disc and a bulge, each of which is associated with a distinct SED, selected from a predefined lookup table. The choice of the galaxy SED is based on specific recipes that are tied to three main galaxy properties: its total stellar mass, its redshift, and its type (star-forming or quiescent).

We modified the original code by adding a third component to account for the AGN emission. For Type 1 AGN (optically unobscured), this component includes the continuum and narrow and broad line emission typical of QSO, while for Type 2 it accounts only for narrow-line emission (i.e., the galaxy stellar continuum is assumed to dominate at all wavelengths). For normal galaxies, this component remains effectively null.

\subsubsection{Type 1 SEDs}\label{subsec:Type1SED}

In order to construct the SEDs of Type 1 AGN in the lightcone, we made use of the parametric model developed by \citet{Temple2021MNRAS.508..737T}. This model allows for the generation of synthetic quasar SEDs over the rest-frame wavelength range $912\AA$ to $3\,{\rm \mu m}$. These synthetic SEDs have been shown to accurately reproduce, to a high degree of accuracy, the observed-frame optical and near-infrared colours of large samples of quasars, over redshifts $0.2 \leq z \leq 7$ and absolute magnitude in the $i$-band $-29 < M_i < -22$.

The observed variety in emission line properties is included in the model through the interpolation between two emission-line templates, which correspond to the observed limits of very strong and very weak emission in terms of the equivalent width of high-ionization ultraviolet lines, such as ${\rm C_{\:IV}}$. Furthermore, observations from quasar spectra show that the equivalent width of strong emission lines is anti-correlated with the intrinsic luminosity of the source. This phenomenon is known as the Baldwin effect \citep{Baldwin1977ApJ...214..679B}. In order to reproduce this phenomenology, the model from \citet{Temple2021MNRAS.508..737T} incorporates a single parameter (\texttt{emline\_type}) that allows for the generation of spectra with different emission line properties. This parameter is related to the absolute magnitude $M_i$ of the quasar by means of the following empirical equation: 

\begin{equation}
{\tt emline\_type} = 0.183 \times (M_i + 27).
\label{eq:Baldwin}
\end{equation}

For this work, we constructed the QSO SED library by changing only this parameter between its minimum and maximum values (-2 to +3), ranging from spectra with weak, highly blueshifted lines to those with strong, symmetric lines.

For each Type 1 AGN, we modified the code EGG to select a SED from the pre-built library of QSO SEDs following Eq.~\ref{eq:Baldwin}. The SED is rescaled to the $L_{{\rm UV}}$ (at $2500 \AA$) that corresponds to that object according to the observed $L_{\rm X} - L_{{\rm UV}}$, as presented in \citet{Bisogni2021A&A...655A.109B}. This relation is parametrised as:

\begin{equation}
\log \left(L_{\rm X}\right)=\gamma \log \left(L_{{\rm UV}}\right)+\beta,
\label{eq:Lx-Luv}
\end{equation}

\noindent where the proxies for the UV and X-ray emissions correspond to the monochromatic rest-frame $2500\AA$ and $2\textrm{ keV}$ luminosities respectively. The authors in \citet{Bisogni2021A&A...655A.109B} find an average value for the slope of the relation of $\gamma = 0.58 \pm 0.06$ up to $z \approx 4.5$ and a dispersion of $\delta = 0.24\,$dex, which we used also for this work. For this, we calculated the monochromatic luminosity at $2\textrm{ keV}$ from $L_{\textrm{X}}$ following the general relation between the total luminosity in the band ($L_{E_1-E_2}$) and at a monochromatic energy ($L_E$), that is:

\begin{equation}
L_E=\frac{(2-\Gamma) E^{1-\Gamma}}{E_2^{2-\Gamma}-E_1^{2-\Gamma}} L_{E_1-E_2},
\label{eq:L2kev}
\end{equation}

\begin{figure}[h]
\centering
\includegraphics[width = \columnwidth]{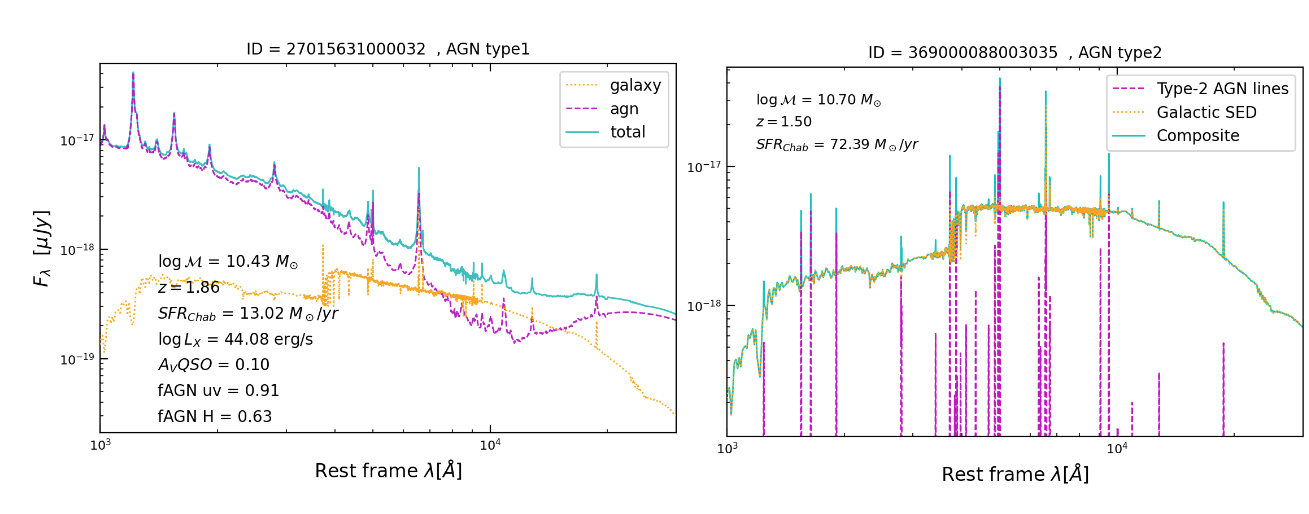}
\caption{Some examples of SEDs of AGN from our lightcone. We show three representative cases: in the upper panel, we show a Type 1 AGN with dominant QSO component in the rest-frame UV band, while in the middle panel, we show a Type 1 AGN with similar QSO and host-galaxy contribution for the same band. The lower panel shows a Type 2 AGN, with a zoom-in subpanel showing the $\rm H_\beta$ and the $[\ion{O}{iii}]\rm\,\lambda\lambda\,5007,4959$ doublet emission lines. The galaxy component (bulge + disc) is shown with an orange dotted line, and is based on  \citet{BC03.2003MNRAS.344.1000B} models. The AGN component, which consists on a full SED template from \citet{Temple2021MNRAS.508..737T} for Type 1 and narrow emission lines for Type 2 is shown with a dashed violet line. The cyan line shows the composed SED. Dust absorption is applied in all cases.  }\label{fig:SEDs}
\end{figure}

\noindent  where in this case $E = E_1 = 2\textrm{ keV}$, $E_2 = 10\textrm{ keV}$, and we used a fixed value for $\Gamma = 1.8$. Therefore, following Eqs.~\ref{eq:Lx-Luv} and~\ref{eq:L2kev} we derive the monochromatic luminosity $L_{2500}$ from the full band luminosity $L_{\textrm{X}}$, and use it to rescale the SED.

We apply the reddening by dust attenuation to the QSO SED using an empirically derived extinction curve presented in \citet{Temple2021MNRAS.508..737T}. This extinction curve is similar to that of the Small Magellanic Cloud for $\lambda \gtrapprox 1700 \AA$, while it increases less rapidly for shorter wavelengths. The $E(B-V)$ of each source is chosen randomly from the distribution presented in Fig. 2 of \citet{Lusso.2013ApJ...777...86L}, which peaks at low $E(B-V)$, with a median value $\langle E(B-V) \rangle = 0.03$.

Finally, the galaxy SED (bulge + disc components) is added to that of the QSO in order to create the composite spectrum. To quantify how much each Type 1 AGN is dominated by the host-galaxy stellar light or by the nuclear emission, we derived the quantity $f_{\rm AGN}$, which is defined as

\begin{equation}
f_{{\rm AGN}_{\rm band}} = \frac{F^{\rm AGN}_{\rm band}}{F_{\rm band}},
\label{eq:fAGN}
\end{equation}

\noindent where $F^{\rm AGN}_{\rm band}$ is the flux in a given band from the AGN component only, and $F_{\rm band}$ is the equivalent for the total spectrum (i.e. disc + bulge + AGN). With this definition, $f_{\rm AGN}$ ranges from 0 to 1, where an object with $f_{\rm AGN} = 0 \, (1)$ would be completely dominated by the galactic (AGN) component, and $f_{\rm AGN} = 0.5$ corresponds to the limit where both the galactic and the nuclear components contribute equally to that specific band. We derived this quantity in two bands: using the rest-frame magnitude from the GALEX FUV filter (which corresponds to the wavelength range where AGN emission typically peaks), and using the observed magnitude $m_H$ (see Table~\ref{tab:filters_colours}).

In Fig.~\ref{fig:SEDs} we show an example of rest-frame SED for both a Type 1 and a Type 2 AGN from the lightcone. We show also in the figure the value of the most relevant physical parameter related to the construction of the SED, such as $\mathcal{M}$, $L_{\rm X}$, SFR and, only for the Type 1 AGN, $f_{\rm AGN_{UV}}$ and $f_{{\rm AGN}_H}$.

\begin{table}[]
\centering
\caption{Filters used in this work.} 
\begin{tabular}{llll}
\hline
Filter             & $\lambda_{\rm ref}$ & $\lambda_{\rm min}$ & $\lambda_{\rm max}$ \\\hline
$I_{{\rm E}}$  & \num{7103}            & \num{5300}       & \num{9318}          \\
$H_{{\rm E}}$  & \num{17649}           & \num{14971}      & \num{20568}         \\
$u$            & \num{3680}            & \num{3206}       & \num{4081}          \\
$z$            & \num{8685}            & \num{8035}       & \num{9375}          \\
GALEX FUV      & \num{1535}            & \num{1340}       & \num{1809}          \\\hline
\end{tabular}

\tablefoot{ All wavelengths are given in $\AA$. Source: SVO Filter Profile Service \citep{SVO.2012ivoa.rept.1015R}.}
\label{tab:filters_colours}
\end{table}

\subsubsection{Type 2 SEDs}\label{subsec:Type2SED}
For objects flagged as Type 2 AGN, typical narrow emission lines from AGN are added to the SED of the bulge component of the host galaxy. The emission lines are modelled with a Gaussian profile, with a total luminosity that was computed using photoionisation models made by \citet{Feltre2016MNRAS.456.3354F}. These models were generated using a standard photoionisation code (\textsc{Cloudy}, \citealp{Ferland.2017RMxAA..53..385F}), and in them, the luminosity of the lines depends on a series of input parameters which describe the physical properties of the region where the narrow lines of AGN are emitted (the narrow-line region, or NLR), and which we explain in detail below:\\

\textit{Gas metallicity: $\log (O/H) + 12$.} The gas-phase oxygen abundance is chosen randomly in the range  $ 8.7 < \log (O/H) + 12 < 9.3$, where we adopted the value $\log (O/H)_\odot + 12 = 8.71$ for the (gas-phase) solar metallicity, as in \citet{Feltre2016MNRAS.456.3354F} and \citet{Gutkin.10.1093/mnras/stw1716} for a corresponding dust-to metal mass ratio $\xi_{d} = 0.3$ (see below). The fact that we are using only models with solar or super-solar metallicities for the NLR of Type 2 AGN is motivated by the fact that these are the models which best sample the region covered by local Seyfert galaxies in standard emission-line diagnostic diagrams \citep{Feltre2016MNRAS.456.3354F}. These values are also consistent with those found in local AGN from observations \citep[e.g.][]{Peluso.2023ApJ...958..147P}.\\

\textit{Ionising spectrum.} The ionising spectrum used in the \textsc{Cloudy} models to represent the accretion disc of the AGN has the shape $S_\nu \propto \nu^\alpha$ for the wavelength range $0.001 \leq \lambda / \mu {\rm m} \leq 0.25$, where the power-law index $\alpha$ is an adjustable parameter. We used only models with either $\alpha = -1.4$ or $\alpha = -1.7$, randomly chosen, which sample the centre of the range modelled by \citet{Feltre2016MNRAS.456.3354F}, i.e. $-1.2$ to $-2$.\\

\textit{Ionisation parameter: $\log U$.} The ionisation parameter is defined as the dimensionless ratio of the number density of H-ionising photons to that of hydrogen. Using a combination of photoionisation models and high-resolution cosmological zoom-in simulations of galaxies, \citet{Hirschmann17.10.1093/mnras/stx2180} found that, at fixed stellar mass, $U$ is one of the main physical parameters driving the cosmic evolution of optical-line ratios. To reproduce this effect in our catalogue, the ionisation parameter is selected randomly (evolving with redshift) within the following ranges\footnote{These values correspond to the volume-averaged ionisation parameter $\langle U \rangle$, as defined in equation 1 of \citet{Hirschmann17.10.1093/mnras/stx2180}. Instead, \citet{Feltre2016MNRAS.456.3354F} used a different definition, namely the ionisation parameter at the Strömgren radius ($U_S$). The conversion between these two quantities is $U_S = \frac{4}{9} \langle U \rangle$. }:

\begin{equation}
\left\{\begin{array}{lll}
-5 < \log U < -3     & {\rm for} & 0 < z < 1 \\
-4 < \log U < -2     & {\rm for} & 1 < z < 2 \\
-3.5 < \log U < -1.5 & {\rm for} & z > 2,
\end{array}\right.
\end{equation}

\noindent where these ranges have been derived from \citet{Hirschmann17.10.1093/mnras/stx2180}  (their fig. 6, central panel). We note that this is the only NLR parameter in our simulation for which we assumed an evolution with redshift, and therefore the redshift evolution of AGN narrow-line ratios is purely linked to that of $\log U.$\\

\textit{Hydrogen number density $n_{\rm H}$}, i.e., volume-averaged hydrogen density of the narrow-line region. Chosen randomly $n_{\rm H} = 10^3$ or $n_{\rm H} = 10^4 \,  {\rm cm}^{-3}$, which are typical gas densities estimated from optical line-doublet analyses of NLRs \citep[see e.g.][]{Osterbrock.2006agna.book.....O, Binette.2024A&A...684A..53B}.\\

\textit{Dust-to-metal mass ratio $\xi_d$,} which accounts for the depletion of metals onto dust grains in the ionised gas. We used only models with $\xi_{d} = 0.3$, which implies assuming that 30 per cent by mass of all heavy elements are in the solid phase \citep{Feltre2016MNRAS.456.3354F, Gutkin.10.1093/mnras/stw1716}. \\

In these models, the intensity of the lines is scaled to the accretion luminosity $L_{\rm acc}$ of the AGN, that is, the luminosity due to the accretion onto the central black hole. Assuming that the bolometric luminosity coming from the AGN, $L_{\rm bol}$, is the sum of $L_{\rm acc}$ and $L_{\rm X}$, and that $L_{\rm bol}$ can be retrieved from the X-ray luminosity with a bolometric correction ($L_{\rm bol} = k_{\rm bol}L_{\rm X}$), $L_{\rm acc}$ can be deduced from the following equation:

\begin{equation}
    L_{\rm acc} = L_{\rm X}(k_{\rm bol} - 1),
\label{eq:Lacc}
\end{equation}

\noindent where again we chose $k_{\rm bol} = 25$ for consistency with the rest of the work.

The emission lines of type 2 AGN are characterised by velocity dispersions which are higher than those of SF galaxies (with FWHM lower than a few hundreds $\textrm{km s}^{-1}$), but lower than those of their type 1 counterparts (with FWHM $\gtrsim 1000 \textrm{ km s}^{-1}$). To model this, we used the results from \citet{Menzel.2016MNRAS.457..110M}, who studied the spectroscopic properties of a sample 2578 X-ray selected AGN in the redshift range $z=[0.02,5.0]$. We modelled the FWHM distribution of the ${\rm H}\beta$ emission line (for ${\rm FWHM}_{{\rm H}\beta} \lesssim 1000 \textrm{ km s}^{-1}$) shown in their Figure 6 as a lognormal distribution centred at $355 \textrm{ km s}^{-1}$, covering the range ${\rm FWHM}\sim [200,1000] \textrm{ km s}^{-1}$ . We assigned randomly the velocity of dispersion of type 2 AGN in our catalogue following this distribution.

Finally, it is worth noting that we did not add an AGN component to the continuum of the host galaxies of type 2 AGN. While at UV and optical wavelengths this approach can be a good approximation for galaxies with strongly attenuated AGN emission, the AGN contribution is expected to dominate in the mid and far IR, even for the most attenuated sources, due to the emission from the dusty torus surrounding the accretion disc. The inclusion of such AGN component is left to future work.

The full list of narrow-region lines added to the spectra of Type 2 AGN is reported in Tab.~\ref{tab:narrow_lines}. We show in Fig.~\ref{fig:SEDs} an example of Type 2 AGN with these lines. In a similar fashion as we did for Type 1 AGN, we quantified the dominance of the Type 2 AGN component with respect to the total (AGN + host galaxy) using the ratio of the emission line flux of $[\ion{O}{iii}]_{\lambda\,5007}$ from the AGN component with respect to the total. This quantity is shown also in Fig.~\ref{fig:SEDs}.

\begin{table}[htbp]
\centering
\caption{Narrow-region lines added to Type 2 AGN spectra.}
\label{tab:narrow_lines}
\begin{tabular}{@{}llll@{}}
\toprule
\textbf{Line} & \textbf{Wavelength} & \textbf{Line} & \textbf{Wavelength } \\
\midrule
	\ion{Ne}{v}                  & 1242.80 &  $\left[\ion{O}{i}\right]$   & 6302.05 \\
	\ion{C}{iv}                  & 1549.86 &  $\left[\ion{O}{i}\right]$   & 6365.54 \\ 
	\ion{He}{ii}                 & 1640.42 &  $\left[\ion{N}{ii}\right]$  & 6549.86 \\
	$\ion{C}{iii}]$              & 1907.71 &  $\left[\ion{N}{ii}\right]$  & 6585.27 \\
	\ion{Mg}{ii}                 & 2795.53 &  $\left[\ion{S}{ii}\right]$  & 6718.32 \\
	\ion{Mg}{ii}                 & 2802.71 &  $\left[\ion{S}{ii}\right]$  & 6732.71 \\
	$\left[\ion{Ne}{v}\right]$   & 3426.85 &  $\left[\ion{S}{iii}\right]$ & 6312.06 \\
	$\left[\ion{O}{ii}\right]$   & 3728.49 &  $\left[\ion{S}{iii}\right]$ & 9071.10 \\
	$\left[\ion{Ne}{iii}\right]$ & 3870.16 &  $\left[\ion{S}{iii}\right]$ & 9533.20 \\
	\ion{He}{ii}                 & 4686.01 &  \multicolumn{2}{l}{Balmer series (${\rm H}\alpha$ to ${\rm H}\eta$)} \\
	$\left[\ion{O}{iii}\right]$  & 4960.30 &  \multicolumn{2}{l}{Paschen series (${\rm Pa}\alpha$ to ${\rm Pa}\eta$)} \\
	$\left[\ion{O}{iii}\right]$  & 5008.24 &   &  \\
\bottomrule
\end{tabular}
\tablefoot{Wavelengths are given in vacuum, in units of $\AA.$}
\end{table}

\section{Validation of the catalogue}\label{sec:validation}
In this section we compare the physical properties of the AGN in our catalogue with data that were not used for its calibration, in order to test the validity of its predictions.

\subsection{AGN fraction}
\begin{figure}[h]
\centering
\includegraphics[width = \columnwidth]{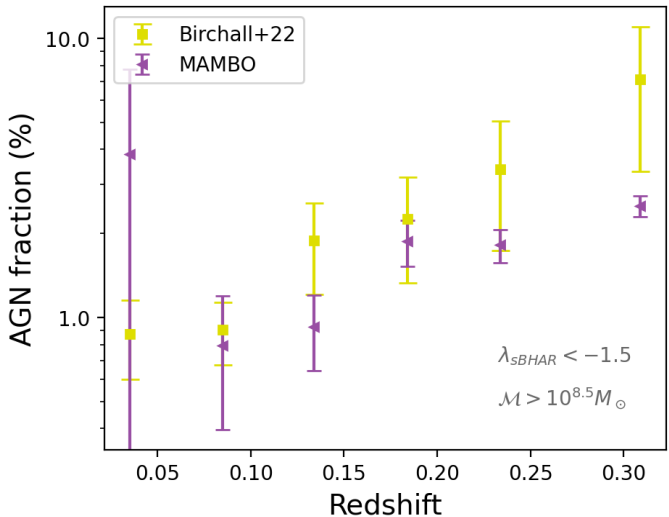}
\caption{Fraction of AGN at $z < 0.33$ from our catalogue (purple triangles), selected with $\log \lambda_{{\rm sBHAR}} < 1.5$  to compare them with the fraction reported in \citet{Birchall_2022MNRAS.510.4556B} (yellow squares). The error bars of the MAMBO AGN fraction show the Poissonian uncertainty.}
\label{fig:Birchall}
\end{figure}

The AGN host galaxy mass function that we used to calibrate the fraction of AGN over the total galaxy population at $z < 2$ \citep{Bongiorno16} is defined for $z > 0.30$ and $\mathcal{M} > 10^{9.5} \, M_\odot$. To test the validity of the extrapolations we did at lower redshifts and stellar masses, we compare our results with different recent works. For example, in Fig.~\ref{fig:Birchall} we compare the fraction of AGN at $z < 0.33$ in different redshift bins with the results from \citet{Birchall_2022MNRAS.510.4556B}, who studied 917 X-ray counterparts of SDSS galaxies. To make our sample as similar as possible to that of \citet{Birchall_2022MNRAS.510.4556B}, we further selected the MAMBO AGN with $\log \lambda_{{\rm sBHAR}} < 1.5$ and $\mathcal{M} > 10^{8.5} \, M_\odot$. We see that for all redshift bins the fractions are roughly consistent within their uncertainities.

Regarding low-mass galaxies, \citet{Latimer2021ApJ...922L..40L} studied a sample of 495 dwarf local galaxies ($\mathcal{M} \leq 3\times10^{9} \, M_\odot$, $z \leq 0.15$) observed by eRosita, and found an upper limit of $1.8\%$ for the AGN fraction. In the same redshift and mass bin our catalogue yields a fraction of $1.4 \pm 0.2\%$, which is roughly compatible with these results. On the other hand, \citet{Mezcua.2024MNRAS.528.5252M} found a much higher AGN fraction (about $20\%$) when studying a sample of dwarf local galaxies from the MaNGA survey using integral-field spectroscopy. We note that the exact AGN fraction at these low stellar mass and redshift regimes is still a matter of debate, and therefore the predictions from our catalogue should be taken with caution here. 


\subsection{X-ray luminosity}

\begin{figure*}[h]
\centering
\includegraphics[width = \textwidth]{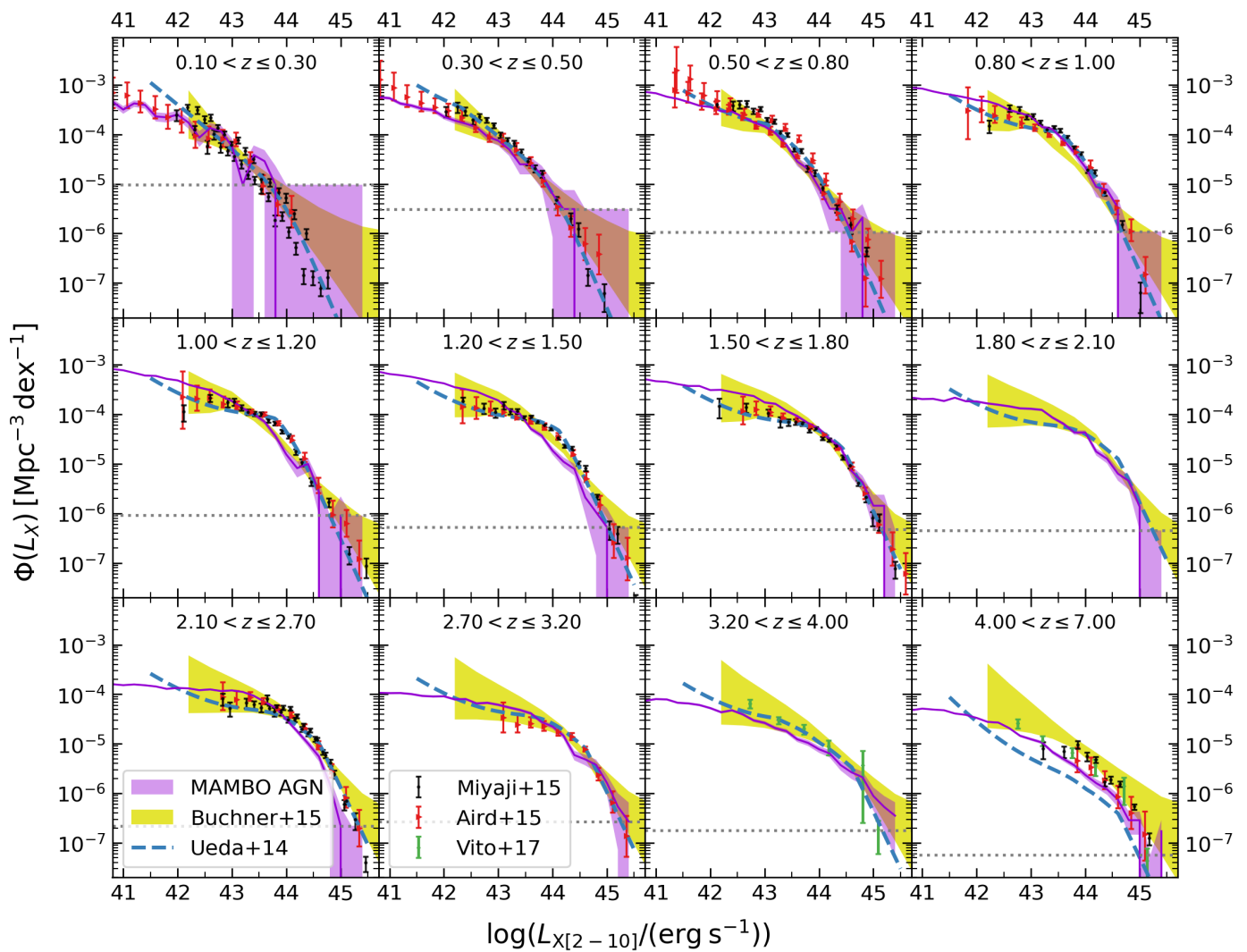}
\caption{Hard X-ray luminosity function of the total population of AGN of our catalogue in different redshift bins, shown by the pruple line. The shaded region represents the Poissonian uncertainty and the dotted horizontal line marks the limiting density from our lightcone (corresponding to 1 object\,$\rm Mpc^{-3}\,dex^{-1}$). For comparison, we show several observed XLFs: \citet{Ueda2014ApJ...786..104U}, \citet{Buchner2015ApJ...802...89B}, \citet{Miyaji.2015ApJ...804..104M}, \citet{XLF.Aird2015.10.1093/mnras/stv1062} and \citet{XLF.Vito.2017.10.1093/mnras/stx2486}.}
\label{fig:xlf}
\end{figure*}


In order to validate the X-ray properties of our catalogue we used two main observables, the X-ray luminosity function and the $F_{{\rm X}}$ cumulative number counts. These comparisons are relevant since they give hints about the purity and completeness of our catalogue in comparison to other X-ray selected catalogues.

Figure~\ref{fig:xlf} shows the hard X-ray luminosity function (XLF) of our mock catalogue compared with several observational luminosity functions from the literature. We note that the XLF by \citet{Miyaji.2015ApJ...804..104M} was used in B16 as an extra observational constraint to determine their HGMF and SARDF, and therefore our catalogue should reproduce it by construction. In Fig.~\ref{fig:xlf} the XLF from \citet{Buchner2015ApJ...802...89B} is scaled by a factor $0.65$ in order to remove the contribution of Compton-Thick (CTK) sources, which were not considered in the work of B16, and therefore are not represented in our catalogue. This factor was chosen because \citet{Buchner2015ApJ...802...89B} found a constant fraction of CTK objects over the total AGN population of about $35\%$ independent of $z$ and $L_{\rm X}$. We observe in general a good agreement between the XLF from our mock and the observed ones at all redshift bins.


\begin{figure}[h]
\centering
\includegraphics[width = \columnwidth]{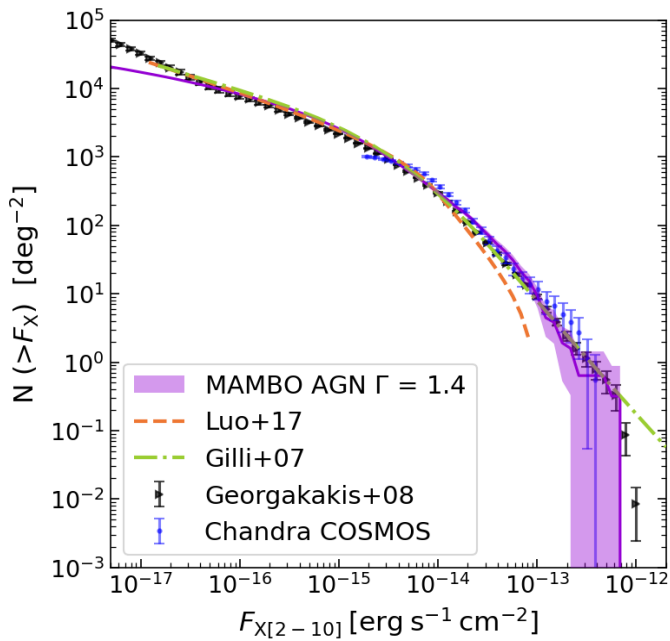}
\caption{Cumulative number counts of X-ray flux in the hard band. The violet shaded area corresponds to the AGN in the MAMBO catalogue, and it shows the uncertainty of our values estimated as the Poissonian error. We show as comparison the number counts reported in different works. References: \citet{Luo.2017ApJS..228....2L}, \citet{Gilli.2007A&A...463...79G}, \citet{Georgakakis.2008MNRAS.388.1205G}, \citet{Marchesi.2016ApJ...830..100M}.}
\label{fig:X_number_counts}
\end{figure}
As a further check, in Fig.~\ref{fig:X_number_counts} we show the cumulative number counts of objects above a given X-ray flux. For this, we estimated the X-ray flux from $L_{\rm X}$ using

\begin{equation}
    F_{{\rm X}} = \frac{L_{\rm X}}{4\pi D_L^2 K(z)},
\end{equation}

\noindent where $D_L$ is the luminosity distance and $K(z)$ is a K-correction of the form

\begin{equation}
    K(z) = (1 + z)^{\Gamma - 2},
\label{eq:K-correction}
\end{equation}

\noindent where $\Gamma$ is the slope of the X-ray spectrum. For the K-correction we assumed a photon index of $\Gamma = 1.4$ which corresponds to the slope of the cosmic X-ray background, and therefore should represent the full population of both obscured and unobscured objects. By comparing the $N(>F_{{\rm X}})$ from our catalogue in Fig.\ref{fig:X_number_counts} to different data from the literature, we see overall a good agreement over a large range of flux until $F_{{\rm X}} \gtrapprox 2\times10^{-17} \, \rm erg\,s^{-1}\,cm^{-2}$, that is, two orders of magnitudes fainter than the sample used to calibrate our methodology.

\subsection{Narrow emission lines}
Diagnostic diagrams are a frequently utilised tool for identifying AGN. By comparing the ratios of various emission lines, we can gain insight into whether star formation, AGN or a composite of both processes dominate in the spectra of a given galaxy. In this section, we employ various optical nebular line diagnostics in order to validate the AGN on our catalogue.

\begin{figure}[h]
\centering
\includegraphics[width = \columnwidth]{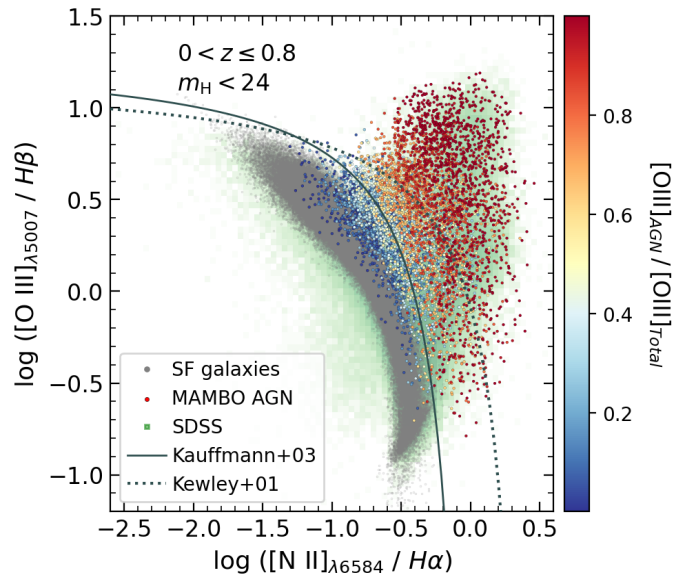}
\caption{BPT diagram for galaxies and AGN from the MAMBO lightcone selected with $z \leq 0.8$ and $m_{H_{\rm E}} < 24$. Galaxies are shown in grey, while AGN are colour-coded from blue to red according to the ratio of the emission line flux of [\ion{O}{iii}] from the AGN component with respect to the total (AGN + host galaxy). Objects from the SDSS DR8 are shown in green in the background, and are traditionally classified as star-forming galaxies if the fall at the left of the \citet{Kauffmann.2003MNRAS.346.1055K}, as AGN if they fall at the right of the \citet{Kewley.2001ApJ...556..121K} line, or as composite if they fall in between these two lines.}
\label{fig:bpt}
\end{figure}

\begin{figure}[h]
\centering
\includegraphics[width = \columnwidth]{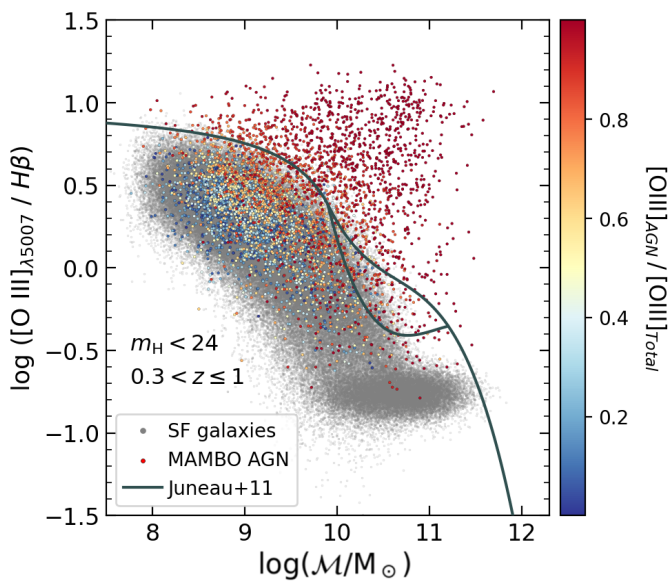}
\caption{Mass-Excitation (MEx) diagram for galaxies and AGN from the MAMBO lightcone selected with $0.3 \leq z \leq 1.0$ and $m_{H_{\rm E}} < 24$. Galaxies are shown in grey, while AGN are colour-coded from blue to red according to the ratio of the emission line flux of [\ion{O}{iii}] from the AGN component with respect to the total (AGN + host galaxy). The solid black lines show the empirical division found in \citet{Juneau.2011ApJ...736..104J} to separate AGN (above the line) from star-forming galaxies (below the line) and composite galaxies (between the two lines). }
\label{fig:mex}
\end{figure}

One of the diagrams used for this purpose is the BPT diagram \citep{BPT.1981PASP...93....5B}, which plots the ratio of the optical lines $[\ion{O}{iii}]_{\lambda5007}$/${\rm H}\beta$ against $[\ion{N}{ii}]_{\lambda6584}$/${\rm H}\alpha$. In this diagram, the \citet{Kauffmann.2003MNRAS.346.1055K} and \citet{Kewley.2001ApJ...556..121K} lines are commonly used to separate between objects dominated by AGN emission (that fall on the top-right of the diagram), those dominated by star-formation (in the bottom-left) and those that have composite spectra (in the central region). 

We show in Fig.~\ref{fig:bpt} an example of such a diagram for the objects of our catalogue, selected with $z \leq 0.8$, since as shown by \citet{Kewley.obs.2013ApJ...774L..10K}, this would be the maximum redshift up to which is safe to use the BPT diagram to separate SF galaxies from AGN. For clarity of visualisation, we show only galaxies and AGN selected with observed magnitude $m_{H} < 24$. The AGN in this figure are colour-coded according to the ratio of the emission line flux of [\ion{O}{iii}] from the AGN component with respect to the total (AGN + host galaxy). We remind the reader that in our mock, the emission line flux in Type 2 AGN is directly proportional to $L_{\rm X}$, and therefore this ratio is also correlated to $L_{\rm X}$. We also show as comparison the observed line ratios from local galaxies from the SDSS catalogue. We see that, in general, the simulated AGN from our mock fall in the regions that correspond to AGN-dominated or composite objects. While there are some that fall in the region of star-forming objects, the great majority of them are AGN with dominant [\ion{O}{iii}] emission from the host galaxy (and most of them have low X-ray luminosities, $L_{\rm X} < 10^{42} \,\rm erg\,s^{-1}$). In fact, different studies have pointed out that the BPT diagram is biased towards more luminous AGN, missing objects with low X-ray \citep{Birchall_2022MNRAS.510.4556B} or optical luminosity \citep{Schawinski.2010ApJ...711..284S}.


An alternative emission line diagnostic to classify AGN is the Mass-Excitation (MEx) diagram \citep{Juneau.2011ApJ...736..104J} which plots $[\ion{O}{iii}]_{\lambda5007}$/${\rm H}\beta$ against $\mathcal{M}$ and was calibrated to separated star-forming galaxies from AGN at $0.3 < z < 1$. Fig.~\ref{fig:mex} shows the MEx diagram for the lighcone presented in this work, where both galaxies and AGN have been selected with $0.3 \leq z \leq 1.0$ and $m_{H_{\rm E}} < 24$. The limit in magnitude allows for a better comparison with the work from \citet{Juneau.2011ApJ...736..104J}, as it removes the low stellar mass tail of our catalogue. We see that all of the AGN which fall in the AGN locus of the diagram have line emission dominated by the AGN component (at least for the $[\ion{O}{iii}]$ line). However, there are some AGN-dominated sources that are classified as MEx-SF galaxies. Similarly to the previous discussion regarding the BPT diagram, the majority of them have $L_{\rm X} < 10^{42}\,\rm erg\,s^{-1}$, while the MEx diagram was validated using AGN selected above this threshold, and therefore, it is difficult to draw clear conclusions from this sub-sample.

\subsection{AGN colours and UVLF}

An important validation for the catalogue if we intend to reproduce the observed AGN population is the luminosity function. In this section we study the redshift evolution of the AGN UV luminosity function (UVLF) at rest-frame wavelength $\lambda = 1450 \AA$, where the majority of UV rest-frame data on AGN is gathered and where Type 1 AGN typically present a peak in their SED (the so-called "big blue bump").

\begin{figure}[h]
\centering
\includegraphics[width = \columnwidth]{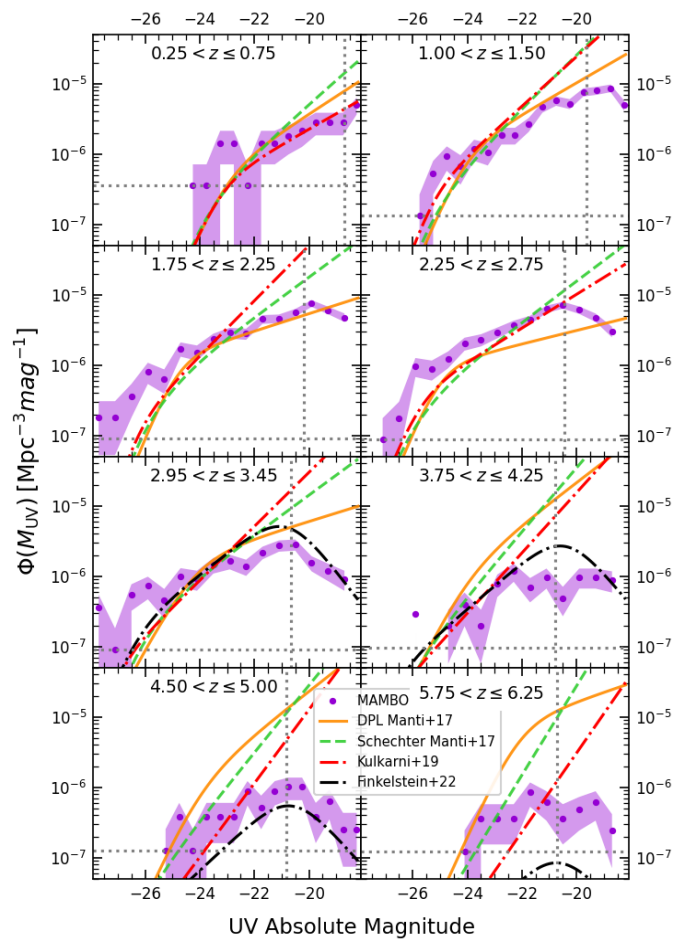}
\caption{UV luminosity function of Type 1 AGN from our catalogue (purple points) compared with different literature LFs in different redshift bins. The purple shaded area shows the uncertainty of our values, estimated as the Poissonian error. In the case of MAMBO, the UV magnitude is computed in the GALEX FUV filter, while for the other cases, it refers to $M_{1500}$. The orange solid and green dashed lines show the parametric LF from \citet{Manti.2017MNRAS.466.1160M}, parameterised as a DPL and a Schechter function respectively. The red and black dash-dotted lines show the DPL 
 parametric LFs from \citet{Kulkarni.2019MNRAS.488.1035K} and \citet{Finkelstein.2022ApJ...938...25F} respectively. The vertical dotted line shows the break magnitude at which the galaxy contribution should start dominating the UVLF \citep{Parsa.2016.10.1093/mnras/stv2857}, while the horizontal dashed line marks the limiting density from our lightcone (corresponding to 1 object$/{\rm Mpc}^{3}/$mag) for each redshift bin. }
\label{fig:UVLF}
\end{figure}

In Fig.~\ref{fig:UVLF} we show the UVLF of the Type 1 AGN from the lightcone presented in this work, in different redshift bins, up to $z = 6$. We have used the absolute rest-frame magnitude from the GALEX FUV filter (see Table~\ref{tab:filters_colours}) as a proxy for $M_{1450}$. We note that the uncertainty shown is only Poissonian and, therefore, constitutes a lower boundary since it doesn't include other systematic effects like selection effects or completeness level, which would increase the uncertainty. 

In Fig.~\ref{fig:UVLF} we also compare our UVLF with different works from the literature. Specifically, we show at all redshift bins the QSO UVLF from \citet{Manti.2017MNRAS.466.1160M}, who parameterised the LF both as a DPL and a Schechter function using a collection of state-of-the-art measurements from z = 0.5 to z = 6.5, and also from \citet{Kulkarni.2019MNRAS.488.1035K}, who used a sample of more than 80 000 colour-selected AGN from redshift $z=0$ to $7.5$ to parameterise the UVLF as DPL evolving with redshift. Both of these works use the absolute monochromatic AB magnitude at a restframe wavelength of 1450 $\AA$ to construct the UVLF. From $z=3$ to $6$ we show also the UVLF from \citet{Finkelstein.2022ApJ...938...25F}, who studied jointly the UVLF of galaxies and QSO and parameterised each population individually with a modified DPL, in order to account for the drop in the faint end of the LF. Additionally, in the figure we show with horizontal dotted line marks the limiting density from our lightcone (corresponding to 1 object$/{\rm Mpc}^{3}/$mag) for each redshift bin. On the other hand, the vertical dotted line shows the break magnitude $M_*$ at $1500 \AA$ of the galaxy UVLF, that is, the magnitude where the galaxy contribution to the ionising background could be relevant, and indeed the galaxy number density higher than the UV/optically selected QSO one. Following \citet{Parsa.2016.10.1093/mnras/stv2857}, \citet{Ricci.2017MNRAS.465.1915R}, this magnitude is calculated as:

\begin{equation}
M_*=(1+z)^{0.206}\left(-17.793+z^{0.762}\right).
\end{equation}

By comparing the UVLF constructed from our catalogue with the ones derived directly from observations, we observe a general agreement up to $z \lesssim 5$, for magnitudes brighter than the break magnitude $M_*$. At fainter magnitudes ($M_{UV}>M_*$) we observe a drop in our QSO UVLF, as expected from the definition of $M_*$ (see above). We note also that at these faint magnitudes there is a big discrepancy also among the observed QSO UVLFs. For example, from $z = 3$ to 5, the faint end of our UVLF is orders of magnitude lower than that of \citet{Manti.2017MNRAS.466.1160M} of \citet{Kulkarni.2019MNRAS.488.1035K}, but agrees quite well with the only-QSO LF of \citet{Finkelstein.2022ApJ...938...25F}.

On the other side, our mock produces an overprediction of the bright end of the UVLF. This can be partially due to the fact we assigned the AGN/galaxy fraction starting from X-ray selected catalogues, which tend to be more complete than UV/optical selections of AGN.

Colour-colour diagrams that use UV to mid-infrared colours can be used to select AGN from a galaxy and AGN sample. Additionally, these selections have the advantage that they can be quickly applied to very large data sets without spectroscopic information. In this section, we use such diagrams to validate the colour properties of the AGN in our mock. For this, we checked different colour-colour diagrams that are known to separate AGN from galaxies. For UV/optical bands, these diagrams are able to select mainly luminous Type 1 AGN, since Type 2 tend to be completely dominated by the host-galaxy emission at these wavelengths.

\begin{figure}[h]
\centering
\includegraphics[width = \columnwidth]{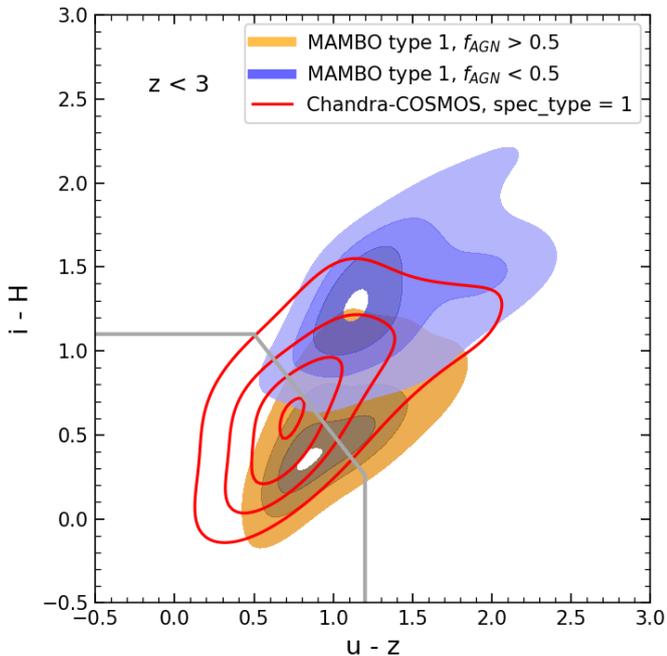}
\caption{Colour-colour diagram to separate galaxies from Type 1 AGN as in Bisigello et al (in prep.). Filled contours show the distribution of Type 1 AGN from the lightcone here presented, separated into those dominated by the AGN component ($f_{\rm AGN} >0.5$, orange contours) or the galaxy component ($f_{\rm AGN} <0.5$, blue contours). The distribution of spectroscopically confirmed Type 1 AGN from the Chandra-COSMOS catalogue \citet{Marchesi.2016ApJ...817...34M} is shown with non-filled red contours. In all cases, the samples have been cut at $z<3$. The contour levels represent iso-density lines, corresponding to the 50th, 75th, 90th, and 99th percentiles of the distribution. The grey line shows the best selection criteria found in Bisigello et al (in prep.) to separate Type 1 AGN from galaxies and Type 2 AGN.}
\label{fig:colour-colour_euclid_lsst}
\end{figure}

In Fig.~\ref{fig:colour-colour_euclid_lsst} we show the ($i - H$) vs ($u - z$) diagram, which was found in \citet{Bisigello.2024arXiv240900175E} to be the best colour selection to separate Type 1 AGN from galaxies using \textit{Euclid} and Rubin/LSST filters. These filters are described in Table~\ref{tab:filters_colours}. The filled contours show the distribution of Type 1 AGN from the lightcone here presented, separated into those dominated by the AGN component in the rest-frame FUV ($f_{{\rm AGN}}\footnote{In this section we use $f_{\rm AGN}$ as shorthand for $f_{{\rm AGN}_{FUV}}$ (Eq.~\ref{eq:fAGN}).} > 0.5$) or the galaxy component ($f_{\rm AGN} < 0.5$). We also show the distribution of spectroscopically confirmed Type 1 AGN from the Chandra-COSMOS catalogue by \citet{Marchesi.2016ApJ...817...34M} as a comparison. In all cases, the distributions have been cut at $z<3$. We observe a good agreement between the Chandra-COSMOS AGN and those with $f_{\rm AGN} >0.5$, while both distributions are clearly separated from that of AGN with $f_{\rm AGN} <0.5$. Also, the MAMBO AGN dominated by the AGN component are in agreement with the best colour by Bisigello et al (in prep.).

\section{Uses for future surveys} \label{sec:uses_future}
This section aims to present some examples of how this catalogue can be used to make predictions that shall aid in the preparation of future large surveys (\textit{Euclid}, Rubin/LSST, Moons, etc.). We focus on two examples, namely the expected number densities of galaxies and AGN in a given photometric band, and the spectroscopic selection of AGN through diagnostic diagrams. For this purpose, we choose the \textit{Euclid} mission as a case study. 

\subsection{Caveats}\label{sec:caveats}
First of all, we would like to warn the reader of a series of caveats to be taken into account when using this catalogue to perform this kind of scientific analysis. Most of them have already been discussed in the text in different sections, but we gather them here for clarity:

The DM lightcone out of which the galaxy and AGN catalogue is constructed contains DM haloes up to $z = 10$. However, the relations we used to construct the galaxy and AGN catalogue are calibrated at lower redshifts. For example, the galaxy SMF is based on observations up to $z = 7.5$, the AGN HGMF up to $z = 2.5$, and the accretion rate distribution up to $z = 4$. Therefore, the extrapolations we did at higher redshifts are to be taken with precaution.

The spectra of Type 2 AGN are constructed by simply adding narrow AGN lines to the continuum of the host galaxy, without adding any nuclear (non-stellar) continuum contribution. This represents a strong assumption which might not hold true for all sources. Furthermore, we separated the AGN in our catalogue into two only categories: Type 1 and 2. In reality, however, we know that such binary classification does not represent the full population of AGN, and AGN are sometimes classified in intermediate classes (1.5, 1.9, etc.).

Likewise, we did not include the IR component from the torus, which usually dominates the continuum emission of Type 1 and 2 AGN for wavelengths redder than $2\,\mu m$.

Throughout this paper we assumed a constant bolometric correction $k_{{\rm bol}} = 25$ ($L_{{\rm bol}} = k_{{\rm bol}}L_{\rm X}$) to ensure self-consistency. However, the bolometric correction is known to correlate with $L_{\rm X}$, and can have a wide range of values covering orders of magnitude \citep{Duras.2020A&A...636A..73D}. 

Finally, we did not include stars in our mock. Some stars, such as cataclysmic variables, have multi-wavelength features that can lead to their misclassification as QSO. Cataclysmic variables appear as point-like sources with QSO-like colours, and often exhibit X-ray emission. Therefore, our mock does not account for contamination from such stellar objects, which is instead present in observations.

We plan to keep developing our pipeline in order to tackle most of these effects.

 \begin{figure}[h]
\centering
\includegraphics[width = \columnwidth]{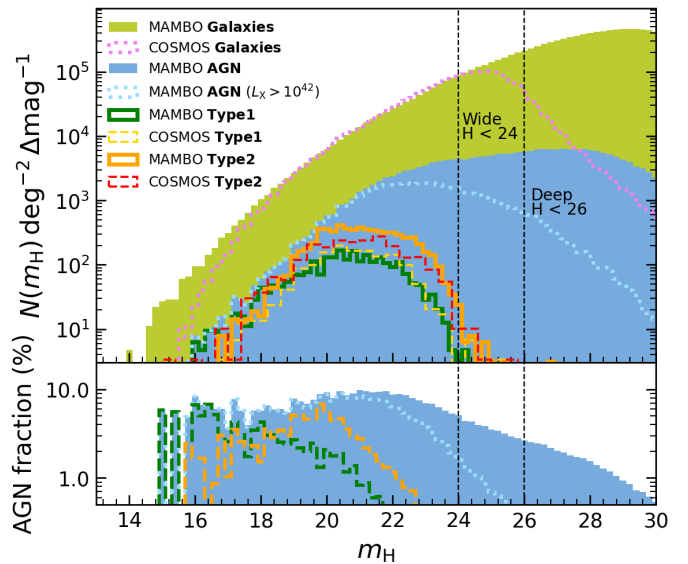}
\caption{Upper panel: Number density counts in the $m_H$ magnitude. Galaxies and AGN from the MAMBO lightcone are represented by green and blue filled histograms respectively. AGN selected with $L_{\rm X} > 10^{42} \,\rm erg\,s^{-1}$ are shown with a cyan dotted line. We also show the density counts for galaxies (pink dotted line) and AGN from the COSMOS catalogue \citep{cosmos15.2016ApJS..224...24L, Marchesi.2016ApJ...830..100M}, where AGN have been selected with $F_{\rm X} > 1.9 \times 10^{-15}\, \rm erg\,s^{-1}\,cm^{-2}$, and separated into Type 1 and 2 based on optical spectroscopic criteria (yellow and red dashed lines). The green and orange solid lines show the number density counts of Type 1 and 2 AGN from our mock after applying the same cut in $F_{{\rm X}}$ as for the COSMOS sample. We show with vertical black dashed lines the limiting magnitude of the EWS and EDS. Lower panel: Fraction of MAMBO AGN as a function of $m_H$. The different lines represent different subpopulations of AGN, as in the upper panel.}
\label{fig:mh_counts}
\end{figure}

\subsection{The \textit{Euclid} Surveys} 
\label{sec:Euclid overview}

The European Space Agency \textit{Euclid} space telescope \citep{EuclidSkyOverview}, successfully launched in July 2023, is equipped with two on-board instruments; the VISible instrument \citep[VIS, ][]{Cropper.2024arXiv240513492E} carries a single broadband optical filter, $I_{\rm E}$. The Near-Infrared Spectrometer and Photometer (NISP, \citealp{NISP.2024arXiv240513493E}), instead, possesses three near-infrared photometric filters: $Y_{\rm E}$, $J_{\rm E}$ and $H_{\rm E}$ (Table~\ref{tab:filters_colours}). The analysis performed in the following subsections is restricted to the $H_{\rm E}$ filter.

Additionally, the NISP instrument is equipped with 4 different low resolution near-infrared grisms (with a spectral resolution of $R=380$ for a 0.5 arcsecond diameter source): 1 blue grism (0.92  to 1.3 $\mu m$), and 3 identical red grisms (1.25  to 1.85 $\mu m$). 

During its 6 year mission, \textit{Euclid} will perform two main surveys: a Wide Survey (EWS), planned to cover an area of \num{14679} square degrees on the sky, and a Deep Survey (EDS) of $53\,{\rm deg}^{2}$ \citep{EuclidSkyOverview}. The limiting magnitude for point sources detected with a minimum S/N of 5 is $m_H = 24$ in the EWS \citep{Scaramella.2022A&A...662A.112E}, and at least two magnitudes deeper in the EDS ($m_H = 26$). During the EWS, the sky will be observed with only one pass of the red grism, which for the H$\alpha$ line translates into a redshift coverage of $0.9 \leq z \leq 1.8$ and a flux limit of $F_{{\rm H}\alpha} > 2\times 10^{-16}\rm erg\,s^{-1}\,cm^{-2}$. Instead, for the EDS the redshift coverage of H$\alpha$ spans to $0.4 \leq z \leq 1.8$ with a limiting flux $F_{{\rm H}\alpha} > 6\times 10^{-17}\rm erg\,s^{-1}\,cm^{-2}$.

\begingroup
\setlength{\tabcolsep}{3pt} 
\renewcommand{\arraystretch}{1.4} 
\begin{table*}[htbp]
\centering
\caption{Surface density and total numbers of galaxies and AGN (full population, Type 1 and Type 2).}
\label{tab:integrated_numbers}
    \begin{tabular}{cccccccc}
    \cline { 2 - 8 }  
    & \multicolumn{2}{c}{Galaxies} & \multicolumn{2}{c}{Type 1 AGN} & \multicolumn{2}{c}{Type 2 AGN}    & All AGN  \\
    \cline { 2 - 8 }
    & Surface density & Total number  & Surface density & Total number & Surface density & Total number & Surface density \\
    &${\rm deg}^{-2}$ &         &${\rm deg}^{-2}$ &          &${\rm deg}^{-2}$ &         &${\rm deg}^{-2}$ \\
    \hline
	 
    \textbf{EWS}   $\,\, m_H < 24$                & $1.4\times10^5$ & $2.1\times10^9$ & $8.3\times10^2$ & $1.2\times10^7$ & $5.5\times10^3$ & $8.0\times10^7$ & $6.3\times10^3$ \\
    \cline { 1 - 1 }
    \textbf{EDS}  $\,\, m_H < 26$                 & $4.0\times10^5$ & $2.0\times10^7$ & $9.4\times10^2$ & $5.0\times10^4$ & $7.4\times10^3$ & $3.9\times10^5$ & $8.4\times10^3$ \\
    \cline { 1 - 1 }
    \textbf{EWS}  $F_{{\rm H}\alpha} > 2\times 10^{-16}$  & \multirow{2}{*}{$2.9\times10^3$} & \multirow{2}{*}{$4.4\times10^7$} &  \multirow{2}{*}{--}  & \multirow{2}{*}{--} & \multirow{2}{*}{$6.1\times10^2$} & \multirow{2}{*}{$8.9\times10^6$} & \multirow{2}{*}{$6.1\times10^2$} \\
    $0.9 \leq z \leq 1.8$ \\
    \cline { 1 - 1 }
    \textbf{EDS}  $F_{{\rm H}\alpha} > 6\times 10^{-17}$  & \multirow{2}{*}{$3.0\times10^4$} & \multirow{2}{*}{$1.5\times10^6$} &  \multirow{2}{*}{--}  & \multirow{2}{*}{--} & \multirow{2}{*}{$2.6\times10^3$} & \multirow{2}{*}{$1.4\times10^5$} & \multirow{2}{*}{$2.6\times10^3$} \\
    $0.4 \leq z \leq 1.8$ \\
	 
    \hline
    \end{tabular}
\tablefoot{Each row corresponds to a different selection, written in the table and explained in greater detail in the text. All AGN are selected with $L_{\textrm{X}} > 10^{42} \,\rm erg\,s^{-1}$. Total numbers assume that the \textit{Euclid} Wide (EWS) and Deep (EDS) surveys will cover areas of $\num{14679}\,{\rm deg}^{2}$ and $53\,{\rm deg}^{2}$ respectively \citep{EuclidSkyOverview}. Fluxes are given in $\rm erg\,s^{-1}\,cm^{-2}$.}
\end{table*}
\endgroup

\subsection{Number densities}
One of the most important predictions that can be done with this type of catalogue is the number densities (number of objects per squared degree) of galaxies and AGN in a given magnitude bin. We show in Fig.~\ref{fig:mh_counts} the number density in the \textit{Euclid} $H$-band for both galaxies and AGN, showing with vertical lines the limiting $m_H$ for the \textit{Euclid} Wide and Deep surveys. We show also for comparison the number density of objects in the COSMOS catalogue. We observe a good agreement between the galaxies from our catalogue and the ones from COSMOS 2015 \citep{cosmos15.2016ApJS..224...24L}. For the AGN sample, we used again the Chandra COSMOS Legacy Spectral Survey catalogue \citep{Marchesi.2016ApJ...830..100M}. We show in Fig.~\ref{fig:mh_counts} the $m_H$ distribution of Type 1 and 2 AGN (classified into these 2 categories by optical spectroscopy), selected with minimum X-ray flux $F_{\rm X} = 1.9 \times 10^{-15}\, \rm erg\,s^{-1}\,cm^{-2}$. After applying the same flux limit to Type 1 and 2 MAMBO AGN, we observe a good agreement between the respective Type 1 and 2 populations in MAMBO and COSMOS. 

 \begin{figure*}[h]
\centering
\includegraphics[width = 0.85\textwidth]{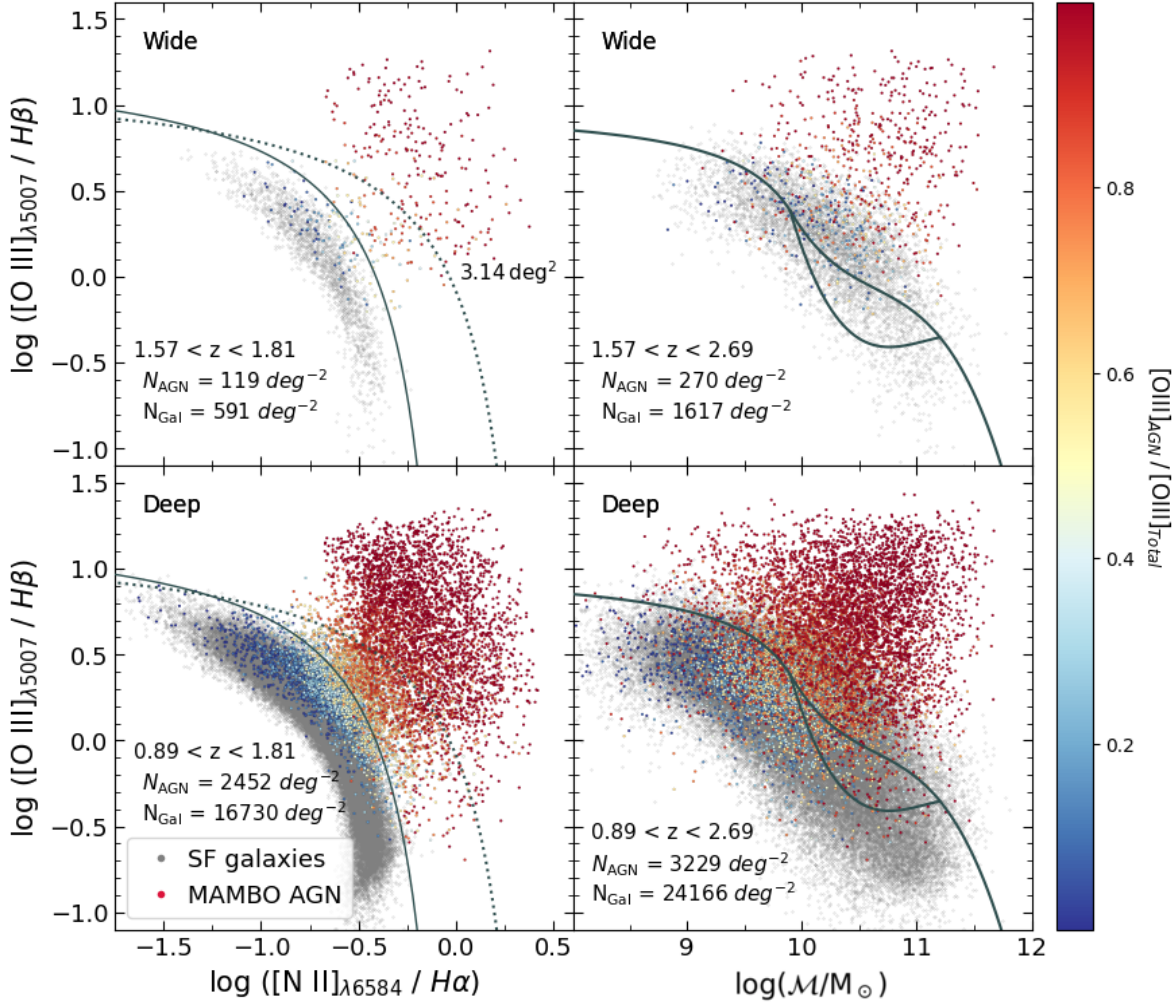}
\caption{Predictions from our lightcone for the BPT (left panels) and MEx diagrams (right panels) as observed by the \textit{Euclid} Wide and Deep surveys. For each panel we have plotted the galaxies (grey dots) and AGN (colour-coded from blue to red) corresponding to the specific redshift range, magnitude limit (in H-band) and ${\rm H}\alpha$ line flux limit of each survey. The surface density of galaxies and AGN in each diagram are given in the figure. See captions of Figs.~\ref{fig:bpt} and~\ref{fig:mex} for further details.}
\label{fig:bpt_mex}
\end{figure*}

Additionally, we show in Table~\ref{tab:integrated_numbers} the expected surface densities and total (integrated) numbers of galaxies and AGN with different selections, in the EWS and the EDS. First, we performed a photometric selection of sources detected above the limiting magnitude of each survey. Furthermore, the numbers shown in Table~\ref{tab:integrated_numbers} consider only AGN selected with intrinsic hard band luminosity $L_{\rm X} > 10^{42} \,\rm erg\,s^{-1}$, to remove possible non-AGN X-ray emitters. We compare these numbers with the ones reported in \citet{Selwood.2024arXiv240518126E}, who performed a similar analysis to the one presented in this work, with the aim of forecasting the expected surface densities of AGN in the \textit{Euclid} Surveys. In tables 4 and 5 of \citet{Selwood.2024arXiv240518126E}, the authors report the surface densities of AGN detectable in the $H_{\rm E}$ band for the \textit{Euclid} Wide and Deep Surveys. For the EWS, the reported surface densities are $4.5 \times 10^3 \, {\rm deg}^{-2}$, $6.8 \times 10^2 \, {\rm deg}^{-2}$ and $1.7 \times 10^3 \, {\rm deg}^{-2}$ for all AGN, Type 1 and Type 2 respectively. The corresponding numbers for the EDS are $2.4 \times 10^3 \, {\rm deg}^{-2}$ , $8.8 \times 10^2 \, {\rm deg}^{-2}$ and $3.5 \times 10^3 \, {\rm deg}^{-2}$. We observe in general a good agreement between these numbers and the ones reported in Table~\ref{tab:integrated_numbers}, especially for the Type 1 AGN, while the numbers differ more for the Type 2.

In Table~\ref{tab:integrated_numbers} we also show the surface densities of sources with ${\rm H}\alpha$ emission line flux above the detection limit of each survey (see Sect.~\ref{sec:Euclid overview}), and only considering sources within the redshift window at which this line will be observed at each survey. For this part, we only considered narrow-line emitters, that is, galaxies and Type 2 AGN, since the values of the emission line fluxes for Type 1 AGN are not included in the current version of the catalogue.

\subsection{Diagnostic diagrams}

In Sect.~\ref{sec:validation} we used two emission line diagnostic diagrams, namely the BPT and the MEx, as validation tools for our lightcone, by studying them at the redshift ranges at which these diagrams have been calibrated. In this section, we studied these same diagrams applying the redshift, magnitude and flux limits corresponding to the \textit{Euclid} Wide and Deep surveys.

Two aspects should be noted in this regard: first, both these diagrams have been calibrated at low redshift ($z \lesssim 1$), while at higher redshifts the physical properties (metallicity, density, ionising radiation, etc.) of the regions where the lines are emitted are expected to change with respect to local conditions. For example, as studied by different works, the position of AGN in the BPT strongly depends on the gas metallicity, and below $Z \sim 0.5 \, Z_\odot$, AGN start to populate the SF region side of the diagram \citep{Groves.2006MNRAS.371.1559G, Kewley.theo.2013ApJ...774..100K, Hirschmann.2019MNRAS.487..333H}. Therefore, these diagrams must be taken more cautiously when used at these redshifts with real data. Secondly, a precise study of the number of AGN that \textit{Euclid} could select with these diagrams (assuming they work at high $z$) would involve using \textit{Euclid}-like spectra with realistic noise and resolution, and official \textit{Euclid} pipelines for the extraction of the line fluxes, which is beyond the scope of this study. Besides, given the spectral resolution of \textit{Euclid}, the $[\ion{N}{ii}]$ and H$\alpha$ lines are likely to be blended in real \textit{Euclid} spectra \citep{Lusso.2024A&A...685A.108E}.

We show in Fig.~\ref{fig:bpt_mex} the BPT and MEx diagrams for the Wide and Deep \textit{Euclid} surveys, applying in each case the corresponding selection in emission line flux and magnitude limit, and redshift range where all relevant lines will be observed. The number density of objects for each case is also given in the figure.

\section{Summary}

We developed an empirical workflow to generate mock catalogues of galaxies and AGN starting from a DM-only simulation . Following \citet{amsdottorato9820} we populated the DM haloes with galaxies by means of a stellar-to-halo mass relation, developed using a subhalo abundance matching technique on observed SMFs. Galaxies were also separated into quiescent or star-forming, following the relative ratio of the blue and red populations in observed SMFs. 

In this paper, we further populated galaxies with AGN following observed host galaxy AGN mass functions at different redshifts and AGN accretion rate distribution functions, which were derived starting from X-ray samples of AGN at $z < 4$ \citep{Bongiorno16, Aird18}. Following \citet{Merloni14.2014MNRAS.437.3550M}, we separated AGN into optically unobscured (Type 1) or obscured (Type 2) and assigned a proper SED to each of them. For this, for Type 1 AGN we used the parametric SED model by \citet{Temple2021MNRAS.508..737T}, which accounts for the continuum emission of Type 1 AGN, as well as their broad and narrow emission lines. For Type 2, instead, we added narrow lines generated using photoionisation models \citep{Feltre2016MNRAS.456.3354F} to the host galaxy stellar continuum.

We tested this workflow by applying it to a $3.14\,{\rm deg}^2$ DM Millennium lightcone up to $z=10$. The result is a mock catalogue of galaxies and AGN with realistic physical properties and observables (such as broadband rest-frame and observed magnitudes and spectral features), complete at least up to magnitude $m_H \sim 28$ and down to stellar mass $\mathcal{M} \sim 10^{7.5} \, M_\odot$. We obtained good agreement between our mock data and state-of-the-art observations such as published $L_{\rm X}$ luminosity functions \citep[e.g. ][]{Buchner2015ApJ...802...89B, XLF.Aird2015.10.1093/mnras/stv1062}, number counts in different NIR to optical bands (in this paper we showed results only on the $H$ band for simplicity), colour-colour diagrams (using $u$, $z$, $i$ and $H$ bands) and emission line diagnostic diagrams (BPT and MEx).

Finally, we demonstrated how this catalogue can be used to make forecasts for future large surveys, using \textit{Euclid} as an example. We computed the expected surface densities of Type 1 and 2 AGN detectable with a given \textit{Euclid} broad filter. We show the results for the $H_{\rm E}$ band, forecasting that \textit{Euclid} will observe about $8.3 \times 10^2$ and $5.5 \times 10^3$ deg$^{-2}$ Type 1 and 2 AGN respectively, selected with $m_{H} < 24$, and $6.1 \times 10^2$ ($2.6 \times 10^3$) deg$^{-2}$ Type 2 AGN with narrow-line $\rm H_\alpha$ emission with flux $F_{{\rm H}\alpha} > 2\times 10^{-16} \, (6\times 10^{-17})\, \rm erg\,s^{-1}\,cm^{-2}$ in the EWS (EDS), finding good agreement with other published forecasts. We also gave examples of the \textit{Euclid} view of narrow-line diagnostic diagrams, which are used to separate local AGN from SF galaxies.

The full workflow is designed to be as computationally efficient as possible so it can be run on a personal computer. In Appendix~\ref{sec:appendix2}, we give more details on the execution time of the main steps of the method.

We plan to update this workflow in the near future in order to tackle the open issues described in Sect.~\ref{sec:caveats}. For example, we plan to revise the AGN fraction and the assignment of the X-ray luminosity using more updated accretion rate distributions \citep[e.g., ][]{Zou.2024ApJ...964..183Z}. Also, we are working on the inclusion of an obscured AGN continuum for Type 2 sources, as well as the emission from a dusty torus, which dominates the IR emission of Type 1 and 2 AGN. Additionally, we plan to investigate the clustering properties of the AGN in our mock catalogues and compare them with observational data as an additional validation step. Although similar studies have been conducted for MAMBO galaxies \citep{amsdottorato9820}, we leave the corresponding analysis for AGN to future work. Finally, we plan to apply this method starting from a bigger DM lightcone.

\section*{Data availability}
The catalogue used for the analysis performed in this paper, together with the necessary code to produce it, are available through the GitHub repository https://github.com/xalolo/MAMBO/tree/main.

\begin{acknowledgements}
We thank the referee for the careful reading and helpful
suggestions that helped to improve the manuscript. We acknowledge the helpful conversations and input from Johan Comparat, Andrea Merloni, Matthew Selwood, James Aird, Matthew J. Temple, Roberto Gilli, Marco Mignoli. XLL acknowledges the "IAC International Scholarships Program". LP and MB acknowledge ASI and Italian Ministry grant "Premiale MITIC 2017".
MB acknowledges the support from INAF Minigrant 2023. 
AF acknowledges the support from the project ``VLT-MOONS" CRAM 1.05.03.07, INAF Large Grant 2022 "The metal circle: a new sharp view of the baryon cycle up to Cosmic Dawn with the latest generation IFU facilities" and INAF Large Grant 2022 ``Dual and binary SMBH in the multi-messenger era". VA acknowledges the support from the INAF Large Grant "AGN $\&$ \textit{Euclid}: a close entanglement" 01.05.23.01.14. IEL acknowledges support from the Cassini Fellowship program at INAF-OAS and the European Union’s Horizon 2020 research and innovation program under Marie Sklodowska-Curie grant agreement No. 860744 “Big Data Applications for Black Hole Evolution Studies” (BiD4BESt).
\end{acknowledgements}

\bibliographystyle{aa}
\bibliography{biblio}

\begin{appendix}

\section{Rejected methodologies}\label{sec:appendix1}

In this appendix, we show different tests which we did before arriving at the final workflow presented in this paper and that motivates some of the choices presented above.

\begin{figure}[h]
\centering
\includegraphics[width = \columnwidth]{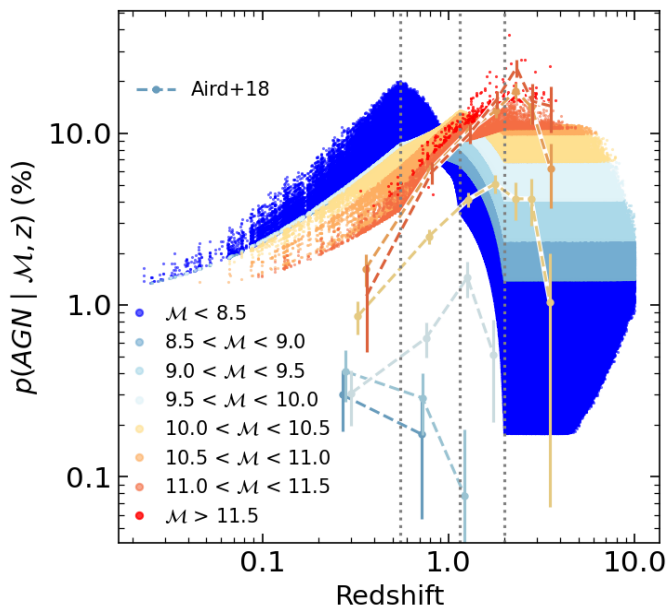}
\caption{Probability of a galaxy to be an AGN as in Fig.~\ref{fig:agn_frac_mam} but using the original Schechter fit from B16 to assign the AGN fraction at $z < 1.15$. The biggest difference with respect to the distribution shown in Fig.~\ref{fig:agn_frac_mam} is in the high density of low-mass AGN at low redshift.}
\label{fig:agn_frac_mam_orig}
\end{figure}

In section~\ref{sec:AGN_frac} we noted that, when using the HGMF from \citet{Bongiorno16}, we modified the slope of the Schechter fit at the two lowest redshift bins. This decision was motivated because the original Schechter fit from B16, when extrapolated to $\mathcal{M} < 10^{9.5}$, predicts a large fraction of low-mass AGN. This can be seen in Fig.~\ref{fig:agn_frac_mam_orig}

\begin{figure*}[h]
\centering
\includegraphics[width = \textwidth]{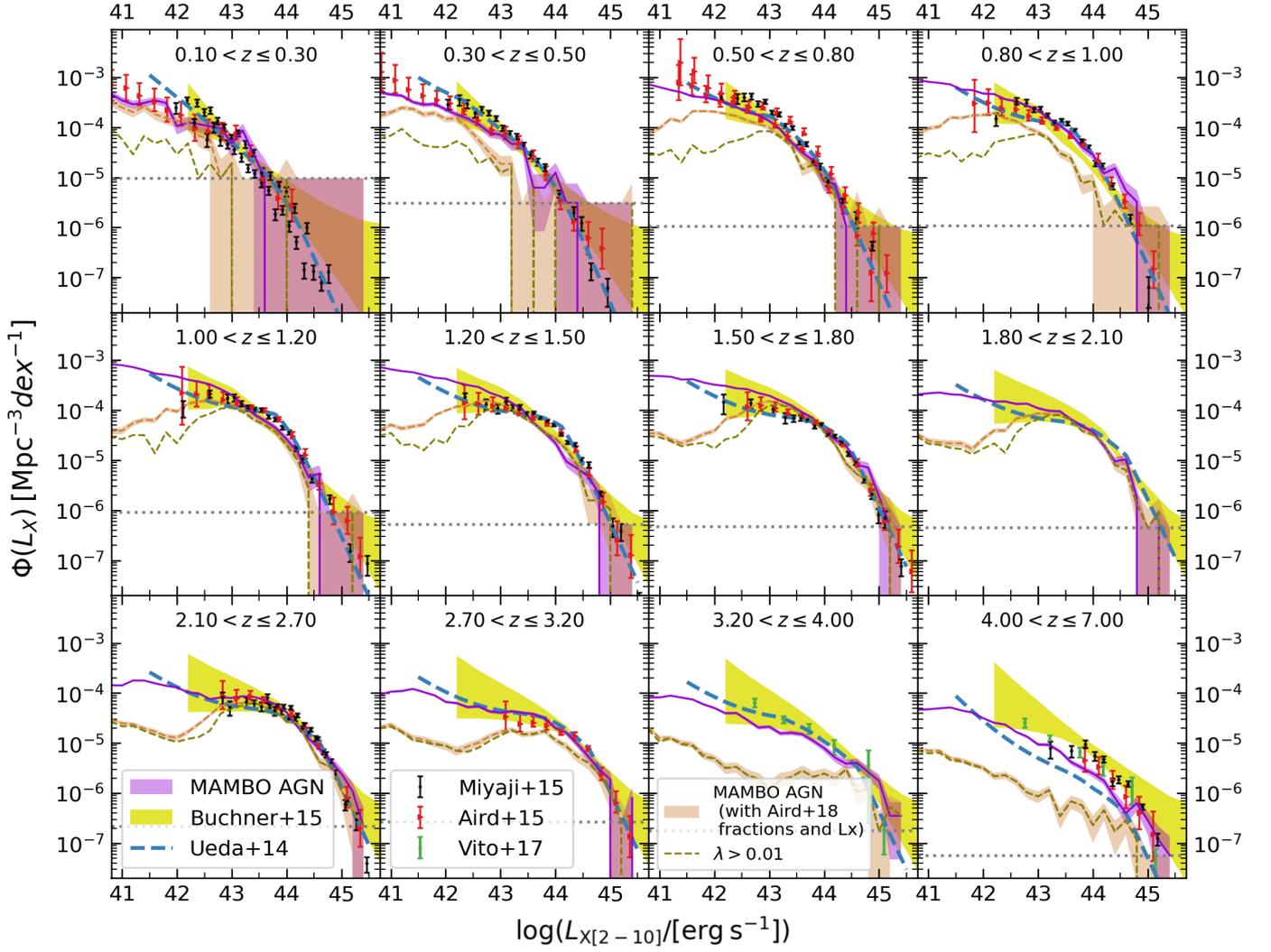}
\caption{Hard X-ray luminosity function, using the $p\left(\log\lambda_{{\rm sBHAR}} \mid \mathcal{M}, z\right)$ distributions from A18 to infer the AGN fraction and $L_{\rm X}$ distribution at all $\mathcal{M}$ and $z$. The dashed brown line corresponds to all the X-ray emitters, while the green dashed line corresponds to the objects selected with $\lambda_{{\rm sBHAR}} > 0.01$. For comparison, the purple solid line shows the XLF from our lightcone using the methodology adopted in this paper, as in Fig.\ref{fig:xlf}. See the caption of Fig.~\ref{fig:xlf} for further details and references to the observed XLFs.}
\label{fig:xlf_aird}
\end{figure*}

\begin{figure*}[h]
\centering
\includegraphics[width = \textwidth]{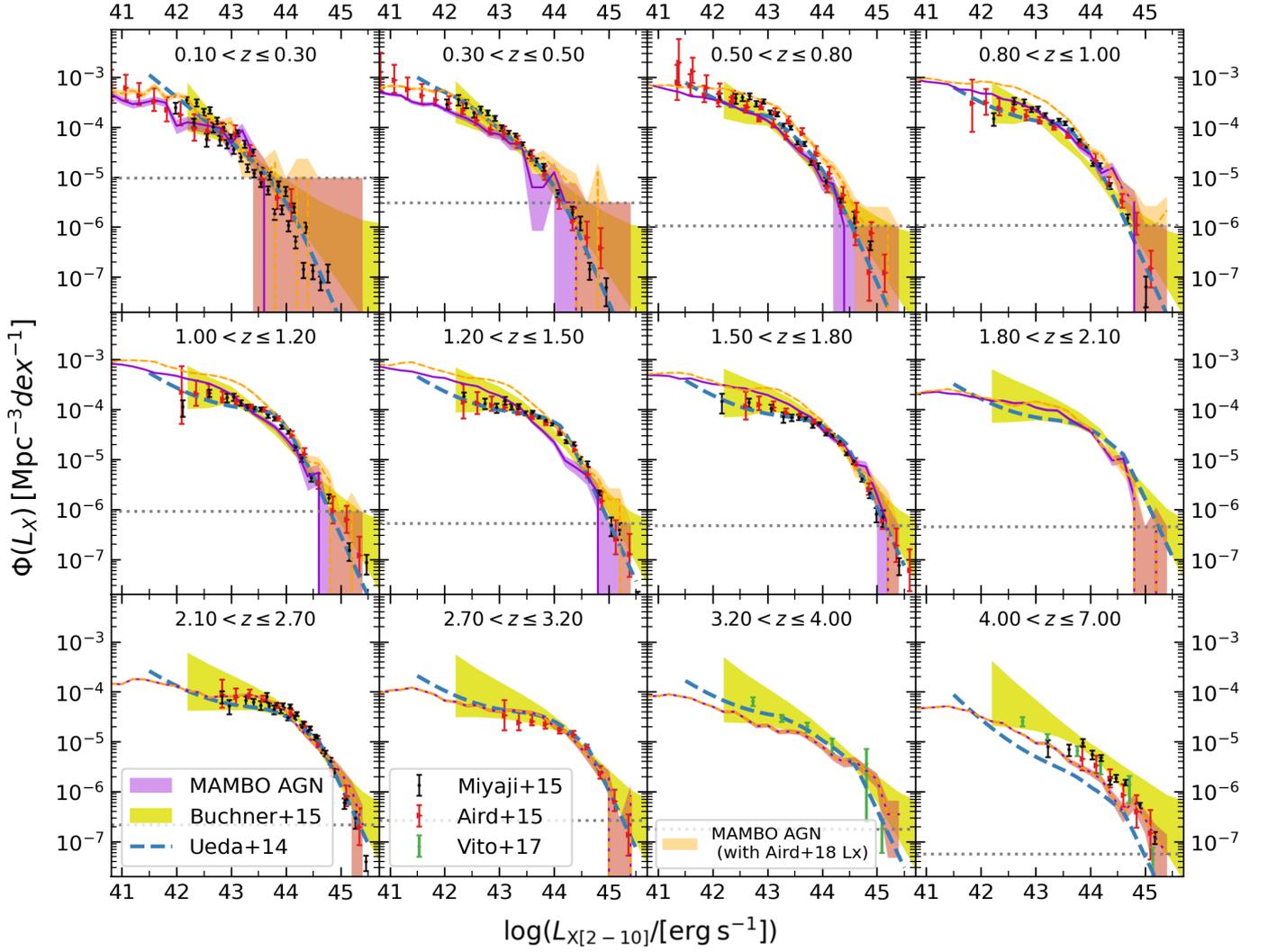}
\caption{Hard X-ray luminosity function, using the AGN fraction inferred from B16 (as in Sect.~\ref{sec:AGN_frac}), but the $p\left(\log\lambda_{{\rm sBHAR}} \mid \mathcal{M}, z\right)$ distributions from A18 to infer $L_{\rm X}$ at all $z$ (dashed orange line). For comparison, the purple solid line shows the XLF from our lightcone using the methodology adopted in this paper, as in Fig.\ref{fig:xlf}. See the caption of Fig.~\ref{fig:xlf} for further details and references to the observed XLFs.}
\label{fig:xlf_aird_onlyLx}
\end{figure*}

In principle, one could avoid doing the steps presented in sections~\ref{sec:AGN_frac} and~\ref{sec:X-ray} and instead derive the AGN fraction and the X-ray distribution in one step, starting from an observed accretion rate distribution. For this, we used the  $p\left(\log\lambda_{{\rm sBHAR}} \mid \mathcal{M}, z\right)$ distributions from A18 at all $\mathcal{M}$ and $z$ and statistically sample random values from those distributions in order to assign $\lambda_{{\rm sBHAR}}$ to every galaxy in our catalogue, and later $L_{\rm X}$ using Eq.~\ref{eq:sbhar}. After this, AGN can be selected as objects above a given threshold in either $L_{\rm X}$ or $\lambda_{{\rm sBHAR}}$.

We show in Fig.~\ref{fig:xlf_aird} the X-ray luminosity function resulting from this methodology, both for AGN selected with $\lambda_{{\rm sBHAR}} > 0.01$ (for consistency with the definition adopted in A18 and in this work), and also for all the X-ray emitters (without any constrain in $\lambda_{{\rm sBHAR}}$). It is visible from this figure that in both cases, the AGN population coming from this methodology is underestimated when comparing it to the observed ones, especially for $z \lesssim 1$ and $z \gtrsim 3$.

Since the $p\left(\log\lambda_{{\rm sBHAR}} \mid \mathcal{M}, z\right)$ distributions from A18 are defined up to $z \leq 4$, we also explored the possibility of using them to infer $L_{\rm X}$ at all $z$, starting from the AGN fraction inferred from B16 (as in Sect.~\ref{sec:AGN_frac}). As it can be seen in Fig.~\ref{fig:xlf_aird_onlyLx}, the results from this methodology tend to overestimate the XLF with respect to observed ones, especially in the range $0.8 \lesssim z \lesssim 1.8$, and therefore we decided to use the approach explained in Sect.~\ref{sec:X-ray}.

\section{Computational time}\label{sec:appendix2}
The most computationally expensive step of the workflow described in this paper is running the modified version of the C++ code EGG. In this step, physical properties and observables are assigned to every galaxy and AGN from the catalogue. The execution time scales mainly with two parameters, namely the number of sources and the number of filters at which one wishes to compute the rest-frame and observed magnitudes. Running EGG on a 2011 iMac with 12 GB of RAM memory and a 2.8 GHz Quad-Core Intel Core i7 processor, with $10^6$ sources and 10 filters both for rest-frame and observed magnitudes took a total time of 36 min 51 s, therefore around 2.2 ms per source. In all our tests, the running time scaled almost linearly with the number of sources.

\end{appendix}

\end{document}